\documentclass[twocolumn,aps,prb,superscriptaddress,floatfix]{revtex4-2}
\usepackage[T1]{fontenc}
\usepackage[utf8]{inputenc}
\usepackage[english]{babel}
\usepackage{color}
\usepackage{float}
\usepackage{braket}
\usepackage{amsmath}
\usepackage{amstext}
\usepackage{amssymb}
\usepackage{graphicx}
\usepackage{bbold}
\usepackage{bm}
\usepackage{dsfont}
\usepackage[unicode=true,pdfusetitle,
bookmarks=true,bookmarksnumbered=false,bookmarksopen=false,
breaklinks=false,pdfborder={0 0 1},backref=false,colorlinks=false]
{hyperref}
\usepackage{ulem}
\newcommand{\angstrom}{\textup{\AA}}
\hypersetup{
	colorlinks,linkcolor=blue,citecolor=blue,urlcolor=blue}
\usepackage{soul}
\makeatletter
\date{}

\makeatother
\setul{0.5ex}{0.3ex}
\definecolor{Blue}{rgb}{0,0.0,1}
\setulcolor{Blue}
\usepackage{babel}

\begin{document}

\title{Controlling electric and magnetic Purcell effects in phosphorene via strain engineering}

\begin{abstract}

We investigate the spontaneous emission lifetime of a quantum emitter near a substrate coated with phosphorene under the influence of uniaxial strain. We consider both electric dipole and magnetic dipole-mediated spontaneous transitions from the excited to the ground state. The modeling of phosphorene is performed by employing a tight-binding model that goes beyond the usual low-energy description. We demonstrate that both electric and magnetic decay rates can be strongly tuned by the application of uniform strain, ranging from a near-total suppression of the Purcell effect
to a remarkable enhancement of more than $1300\%$ due to the high flexibility associated with the puckered lattice structure of phosphorene. We also unveil the use of strain as a mechanism to tailor the most probable decay pathways of the emitted quanta. Our results show that uniaxially strained phosphorene is an efficient and versatile material platform for the active control of light-matter interactions thanks to its extraordinary optomechanical properties. 

\end{abstract}

\author{P. P. Abrantes}
\email{ppabrantes91@gmail.com}
\affiliation{Departamento de Física, Universidade Federal de São Carlos, Rod. Washington Luís, km 235 - SP-310, 13565-905, São Carlos, São Paulo, Brazil}

\author{W. J. M. Kort-Kamp}
\affiliation{Theoretical Division, Los Alamos National Laboratory, MS B262, Los Alamos, New Mexico 87545, USA}

\author{F. S. S. Rosa}
\affiliation{Instituto de Física, Universidade Federal do Rio de Janeiro, Caixa Postal 68528, Rio de Janeiro 21941-972, RJ, Brazil}

\author{C. Farina}
\affiliation{Instituto de Física, Universidade Federal do Rio de Janeiro, Caixa Postal 68528, Rio de Janeiro, 21941-972, RJ, Brazil}

\author{F. A. Pinheiro}
\affiliation{Instituto de Física, Universidade Federal do Rio de Janeiro, Caixa Postal 68528, Rio de Janeiro 21941-972, RJ, Brazil}

\author{Tarik P. Cysne}
\email{tarik.cysne@gmail.com}
\affiliation{Instituto de F\'\i sica, Universidade Federal Fluminense, 24210-346, Niter\'oi RJ, Brazil}

\maketitle 

\section{Introduction \label{sec1}} 

In a pioneering work, E. M. Purcell demonstrated that the surrounding environment could drastically modify the spontaneous emission (SE) rate of an excited quantum system \cite{Purcell1946}. This effect occurs due to the modification of the local electromagnetic density of states and, consequently, the number of available decay channels for the deexcitation of the emitter. The engineering of the SE via the Purcell effect is an accessible tool for probing the optical density of states, leading to a plethora of applications that run from the design of efficient scintillators \cite{Ye-Bizarri-Scintillator-2022} and light-emitting diodes \cite{Kim-Jung-Park-2021, Huang-Chen-Yang-OptExpress-2022} to single-photon sources \cite{Kaupp-Hunger-PRApplied-2016, Jeantet-Voisin-PRL-2016}. The study of the Purcell effect remains an active topic in nanophotonics and has been investigated for emitters near structures of distinct geometries and materials \cite{Blanco-GarciadeAbajo-2004, Rosa-Farina-2008, Biehs-Greffet-2011, Vladimirova-adkov-2012, Kort-Kamp-Farina-2013, Klimov-ACSnano-2015, Klimov-NatMat-2019, Lodahl-Nature-2004, Lodahl-RPM-2015, vanDriel-PRL-2005}.

Quantum emitters are confined systems with discrete electronic spectra subjected to radiative optical transitions. They can either be atoms, molecules, nanoparticles, or even quantum dots. For most quantum emitters, the decay from an excited state to the ground one occurs via the electric dipole (ED) transition \cite{Novotny-book}. There exist, for example, a variety of quantum dots that emit via ED transitions in wavelengths ranging from $0.3$ to $4.1$ $\mu$m \cite{Review-QuantumDots}. Nevertheless, the SE may also occur due to magnetic dipole (MD) transitions \cite{Lodahl-RPM-2015}. Most often, the MD contribution to the SE is weaker than the ED one by a factor of $\alpha=1/137$ \cite{Novotny-book}, so the electric Purcell effect has been usually much more investigated in photonics than its magnetic counterpart. However, recent progress in nanofabrication techniques has allowed for the design of new nanostructures that enhances the MD contribution in relation to the SE \cite{Hussein-Neshev-OLett-2015, THz-Mag-Purcell}. In addition, the SE of rare-earth ions \cite{Alu-magneticSE-Review} and some suitably designed quantum dots \cite{Feng-QD-Lambda-500nm} can also be dominated by MD transitions. Depending on the emitter, the wavelength of the MD transition may vary from $0.5$ to $500$ $\mu$m \cite{Feng-QD-Lambda-500nm, THz-Mag-Purcell}. Recent studies on the magnetic Purcell effect include emitters close to dielectric nanostructures \cite{MPF-Silicon-Nanostructures}, antiferromagnets \cite{Ferreira-Peres-EPL-2019}, and parity-time symmetric potentials \cite{Alaeian-Dionne-PRB-2015}, but its full potential for applications is still unexplored.

The advent of two-dimensional (2D) materials, triggered by the synthesis of graphene nearly two decades ago, has unlocked a new venue in tailoring light-matter interactions down to the nanoscale. In contrast to the usual three-dimensional materials used in photonics, 2D materials possess an electronic structure that can be highly modified by external stimuli with weak or moderate intensities, enabling unprecedented control of light-matter interactions. For instance, the possibility of applying electromagnetic fields to control Casimir and Casimir-Polder interactions on graphene and graphene-family materials has been theoretically explored \cite{Cysne-CasimirPolder-PRA-2014, Silvestre-QR-PRA-2019, Abrantes-QR-PRB-2021, Rodriguez-Lopez-NatCommun-2017, Muniz-Farina-Kort-Kamp-2021}. Similar studies on the Purcell effect \cite{Kort-Kamp-Amorim-FresnelCoefficients}, near-field radiative heat transfer \cite{NFRHTGraphene-PRApp, NFRHTGraphene-PRB}, photonic spin Hall effect \cite{Kort-Kamp-PRL-2017, MShah-JPD-AppPhys}, and resonance energy transfer \cite{Abrantes-RET-2021} have also been performed and, despite the great level of tunability predicted in all these cases, the application of strong external electromagnetic fields may present practical difficulties. Furthermore, 2D materials are experimentally used in nanophotonics \cite{Low-Martin-Moreno-NatureMaterials, Reserbat-Plantey-ACSPhotonics-2021, Liu-Mohideen-PRL-2021, Guest-NanoLett-2018}, prompting the search for novel methods to control their interaction with light.

Phosphorene is a monolayer of black phosphorus, first synthesized in 2014 \cite{Phosphorene-First-Syntesis, Phosphorene-Second-Syntesis}. This atomically thin material has emerged as an appealing platform for application in optics, among other reasons, due to its anisotropic band structure and direct electronic energy gap \cite{Review-Light-Matter-Phosphorene, Rudenko-Katsnelson-Ph-NoStrain, Rodin-Carvalho-CastroNeto-Ph-NoStrain}. Indeed, it was shown that this anisotropy may cause non-trivial changes in the sign of the Casimir-Lifshitz torque \cite{Casimir-Torque-Phosphorene}. Some studies on the ED SE close to phosphorene have also been carried out, analyzing the behavior of its electronic spectra with layer stacking and twisting \cite{Phosphorene-SE-Twisting, Phosphorene-SE-NLayer, Phosphorene-SE-Bilayer, Phosphorene-SE-PRAplied-2019}. In contrast to other 2D materials, the puckered lattice of phosphorene makes its electronic structure very sensible to strain \cite{Peeters-StrainPhosphorene-Model, Midtvedt-Lewenkopf-Croy-JPCM, Midtvedt-Lewenkopf-Croy-2DMat}, and its flexibility allows for sustaining high-strain levels up to $30\%$ \cite{Flexibility-Phosphorene-1, Flexibility-Phosphorene-2}. When subjected to uniaxial strain, which is usually implemented in experiments \cite{Exp-Strain-Ph-1, Exp-Strain-Ph-2}, the energy band gap in phosphorene and the Fermi velocity of the carriers are altered, which modifies the anisotropic character of the material and results in a modification of its optical response.

By means of a more sophisticated tight-binding model that goes beyond the low-energy description commonly used in the framework of nanophotonics to model phosphorene layers \cite{Phosphorene-SE-Twisting, Phosphorene-SE-NLayer, Phosphorene-SE-Bilayer}, we are able to describe the modifications in the material properties due to the application of a uniform strain field. Indeed, we demonstrate that this methodological progress, when applied in the context of nanophotonics, is able to not only successfully describe the optomechanical properties of phosphorene but also unveil unknown optical functionalities so far. Based on such a model, we demonstrate that uniaxially strained phosphorene may affect the SE of electric and magnetic dipole emitters, leading to a remarkable suppression of almost $100\%$ and enhancements of more than $1300\%$ of the Purcell effect. We discuss the situations in which the dipole moment is aligned parallel to the $x$ (armchair), $y$ (zigzag), and $z$ (perpendicular) directions. We show that the intrinsic anisotropy of the phosphorene lattice implies the dependence of the decay rate on the orientation of the electric and magnetic dipoles. Finally, our findings attest that strain can be employed to tailor the probabilities associated with the different decay channels into which the photon can be emitted, demonstrating the impact of the extraordinary optomechanical properties of phosphorene in light emission engineering.



\section{Theoretical model and Results \label{sec1,5}}

We use the tight-binding model for phosphorene developed in Refs. \cite{Rudenko-Katsnelson-Ph-NoStrain, Rodin-Carvalho-CastroNeto-Ph-NoStrain}. This model has been successfully applied in the context of condensed matter physics to describe many of phosphorene's remarkable properties, such as its topological characteristics \cite{Peeters-StrainPhosphorene-Model}, the anisotropic nature of its optical response \cite{Alidoust-Akola-PRB-2021, Yan-Zhang-Wang-Zhang-Optcond}, the quantum transport properties in the presence of disorder \cite{Li-Peeters-2018}, and its mesoscopic physics \cite{Li-Peeters-2017}. Using Harrison's prescription, one can also include the effect of a uniform strain field in the model \cite{Peeters-StrainPhosphorene-Model}. Previous studies on phosphorene applied to nanophotonics used a low-energy description \cite{Faria-Junior-Low-energyModel} simply including a direction-dependent Fermi velocity \cite{Phosphorene-SE-Twisting, Phosphorene-SE-NLayer, Phosphorene-SE-Bilayer}, which captures the phosphorene's anisotropic optical nature. Nevertheless, these models are insufficient to explore phosphorene's strain engineering, one of the prominent characteristics of the material. The tight-binding model for strained phosphorene is reviewed in Appendix \ref{AppA}. As we discuss in the following, the application of this tight-biding model allows for a successful description of the optomechanical properties of phosphorene and unveils the unique quantum emission functionalities that can be harnessed by the presence of strain.

The optical conductivity of strained phosphorene monolayer can be computed from the tight-binding Hamiltonian [Eq. (\ref{HP2})], employing linear response theory \cite{Yan-Zhang-Wang-Zhang-Optcond, Low-Rodin-Optcond}. Here, we neglect spatial dispersion, which is supported by previous numerical calculations using different 2D materials that showed that this approximation accurately describes the Purcell effect for the distance scales we are interested in this work \cite{Non-Local_x_Local-PF-Abajo}. Within these assumptions, one can write the constitutive equation $\bm{J} (\bm{r}, \omega) = \overleftrightarrow{\bm{\sigma}} (\omega, \epsilon_{\mu}) \cdot \bm{E} (\bm{r}, \omega)$, where $\bm{E} (\bm{r}, \omega)$ is the amplitude of the oscillating electric field, $\bm{J} (\bm{r}, \omega)$ is the amplitude of the induced oscillating charge current, and
\begin{eqnarray}
\overleftrightarrow{\bm{\sigma}} (\omega, \epsilon_{\mu})=\begin{bmatrix}
\sigma_{xx}(\omega, \epsilon_{\mu}) & 0  \\
0 & \sigma_{yy}(\omega, \epsilon_{\mu}) 
\end{bmatrix}
\label{TensorSigma}
\end{eqnarray}

\noindent is the optical conductivity tensor of strained phosphorene. In this expression, $\epsilon_{\mu}$ ($\mu= x, y, z$) is the uniform strain in phosphorene applied along the $\mu$ direction. In Appendix \ref{AppB}, we compute the optical conductivity of strained phosphorene in different situations.

\subsection{Electric dipole emission \label{sec2}} 

\begin{figure}[b!]
    \centering
    \includegraphics[width=0.65\linewidth]{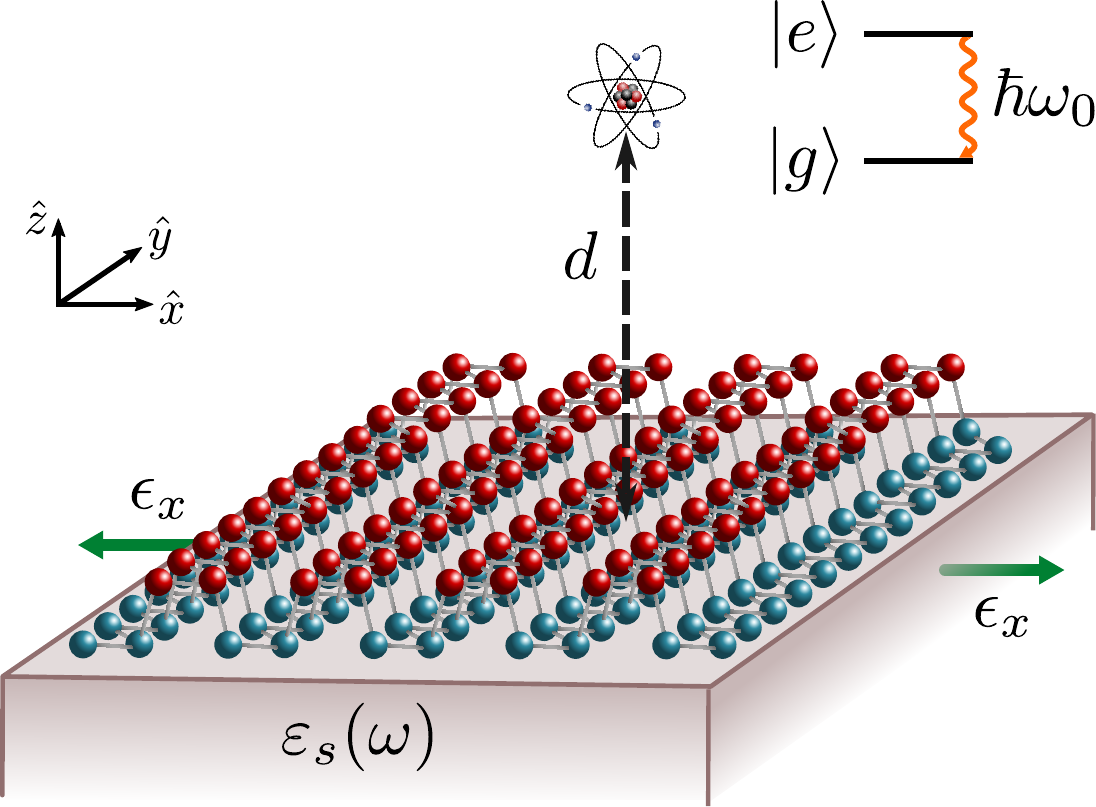}    
    \caption{Quantum emitter at distance $d$ above a phosphorene sheet grown on top of a substrate with permittivity $\varepsilon_s(\omega)$. The phosphorene sheet is under uniaxial strain (applied along the $x$ direction, in this figure) controlled by the substrate.}
    \label{Fig-Boneco}
\end{figure}

\begin{figure*}[t!]
	\centering
	\includegraphics[width=0.9\linewidth]{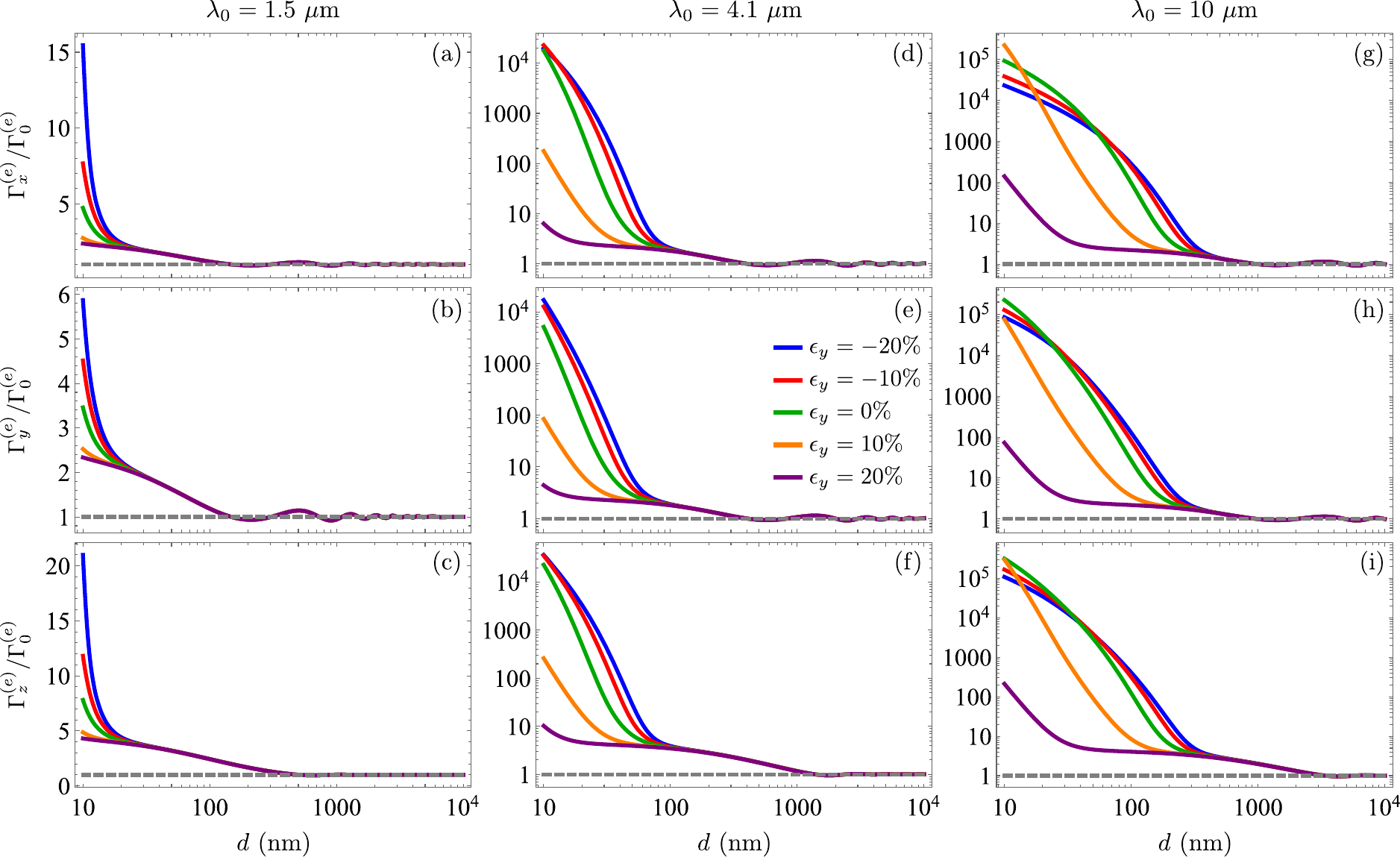}    
	\caption{Electric dipole Purcell factors as functions of the distance between the emitter and the phosphorene/SiC substrate. We considered the uniaxial strain in the $y$ direction with intensities $\epsilon_y=-20, -10, 0, 10, 20 \%$. The emitter's transition wavelengths are (a)-(c) $1.5$ $\mu$m, (d)-(f) $4.1$ $\mu$m, and (g)-(i) $10$ $\mu$m.}
	\label{FigPF-Ele-Sy}
\end{figure*}

We consider the system depicted in Fig.~\ref{Fig-Boneco}. The half space $z<0$ is composed of a homogeneous, isotropic, and nonmagnetic dielectric with permittivity $\varepsilon_{\rm s} (\omega)$. On top of this substrate ($z=0$), a phosphorene sheet is placed. The substrate permits the mechanical application of uniaxial strain in the phosphorene layer. We assume the upper medium $z>0$ to be vacuum, and an excited quantum emitter is located at $\bm{r}_0 = (0,0,d)$. 

We first consider the quantum emitter as a two-level system dominated by an ED transition between the excited $\ket{e}$ and ground $\ket{g}$ states with energy difference $E_e-E_g=\hbar \omega_0=\hbar k_0 c$. The electric Purcell factor (PF) is the modification in the SE rate due to the presence of neighboring objects and can be written as \cite{Szilard_SE_VO2}
\begin{equation}
	\frac{\Gamma^{(e)} (\bm{r})}{\Gamma_0^{(e)}} = \frac{6 \pi c}{\omega_0} \,{\rm Im} \left[ \hat{p} \cdot \mathds{G}^{(e)} (\bm{r},\bm{r}, \omega_0) \cdot \hat{p} \right] ,
\label{EletricPF}
\end{equation}

\noindent where $\Gamma_0^{(e)} = |\bm{p}|^2 \omega_0^3/3 \pi \hbar \varepsilon_0 c^3$ is the free space SE rate of an ED emitter, $\bm{p}$ is the emitter's transition ED moment, $\hat{p} = \bm{p}/|\bm{p}|$, and $\mathds{G}^{(e)} (\bm{r},\bm{r}', \omega)$ is the electric dyadic Green function of the system. One can evaluate the PF writing $\mathds{G}^{(e)} (\bm{r},\bm{r}, \omega)$ in terms of the diagonal part of the reflection matrices \cite{Kort-Kamp-Amorim-FresnelCoefficients}. With the knowledge of the optical conductivity of phosphorene and the electric permittivity of the substrate, one can calculate the desired reflection coefficients by solving the Maxwell equations with the appropriate boundary conditions (see Appendix \ref{AppC}). The expressions of the electric PFs $\Gamma^{(e)}_{x}/\Gamma_0^{(e)}$, $\Gamma^{(e)}_{y}/\Gamma_0^{(e)}$, and $\Gamma^{(e)}_{z}/\Gamma_0^{(e)}$ for the cases of transition ED moments parallel to the $x$ (armchair), $y$ (zigzag), and $z$ (perpendicular) directions, respectively, can be cast as \cite{Purcell-Phosphorene-OptExpress-Sikder}
%
\begin{align}
	\frac{\Gamma^{(e)}_{x}}{\Gamma_0^{(e)}} &= 1 + \frac{3}{4\pi k_0} \, {\rm Im} \left[ i \int d^2 \bm{k}_{\parallel} \frac{e^{2 i \sqrt{k_0^2 - k^2_{\parallel}} d}}{ k^2_{\parallel} \sqrt{k_0^2 - k^2_{\parallel}}} \right. \nonumber \\ &\times \left.\left( k^2_y r_{ss} - \frac{k^2_x (k_0^2 - k^2_{\parallel})}{k^2_0} r_{pp} \right) \right],
\label{gammaElecXX} \\
	\frac{\Gamma^{(e)}_{y}}{\Gamma_0^{(e)}} &= 1 + \frac{3}{4\pi k_0} \, {\rm Im} \left[ i \int d^2 \bm{k}_{\parallel} \frac{e^{2 i \sqrt{k_0^2 - k^2_{\parallel}} d}}{k^2_{\parallel} \sqrt{k_0^2 - k^2_{\parallel}}} \right. \nonumber \\ &\times \left. \left( k^2_x r_{ss} - \frac{k^2_y (k_0^2 - k^2_{\parallel})}{k^2_0} r_{pp} \right) \right],
\label{gammaElecYY} \\
	\frac{\Gamma^{(e)}_{z}}{\Gamma_0^{(e)}} &= 1 +  \frac{3}{4\pi k_0^3} \, {\rm Im} \left[ i \int d^2 \bm{k}_{\parallel} \frac{k^2_{\parallel} \, e^{2 i \sqrt{k_0^2 - k^2_{\parallel}} d}}{\sqrt{k_0^2 - k^2_{\parallel}}} r_{pp} \right], 
\label{gammaElecZZ}
\end{align}
where $r_{ss}$ and $r_{pp}$ are diagonal reflection coefficients (see Appendix \ref{AppC}) and $k_{\parallel}=|\bm{k}_{\parallel}|=|k_x\hat{x}+k_y\hat{y}|$. Due to the anisotropic nature of phosphorene, we obtain $\Gamma^{(e)}_{x} \neq \Gamma^{(e)}_{y}$. Throughout this paper, we consider a silicon carbide (SiC) substrate and, in all results of the main text, we set the Fermi energy of phosphorene at $E_{\rm F}=0.7$ eV. The control of the carriers density to keep the Fermi energy fixed can be done by tuning the back-gate voltage \cite{Das-Roelofs-ACS-Nano-2014}.

In Fig.~\ref{FigPF-Ele-Sy}, we show the PFs as functions of the distance $d$ between the emitter and the phosphorene/SiC medium for different values of uniaxial strain $\epsilon_y~=~-20, -10, 0, 10, 20 \%$. We consider emitters with ED transitions at three distinct wavelengths $\lambda_0 = 2\pi c/\omega_0$, to wit, $1.5$ $\mu$m, $4.1$~$\mu$m, and $10$ $\mu$m, the first two values lying in the near to mid-IR range reached by a wide variety of quantum dots \cite{Review-QuantumDots}. Emitters with longer wavelengths have already been experimentally explored in the context of SE \cite{Hulet-Kleppner-dipoleTHz-PRL-1985}. Comparing the results corresponding to relaxed phosphorene sheets, one can see that the longer the transition wavelengths, the more pronounced the changes in the SE rates are, with the PFs reaching values in excess of $10^5$ when $d = 10$ nm. When strain comes into play, the PFs may be dramatically modified, particularly at small distances. As discussed in Appendix \ref{AppB}, the compressive uniaxial strain ($\epsilon_y<0$) enhances the Drude weight and, consequently, the intraband contribution to the optical conductivity. The opposite occurs in the case of tensile strain ($\epsilon_y>0$), which decreases the Drude weight and the intraband contribution. In most frequency ranges, the interband contribution presents the same behavior. It should be noticed that, for $\lambda_0 = 1.5$~$\mu$m and $\lambda_0 = 4.1$ $\mu$m, these patterns with $\epsilon_y$ are also followed by the PFs: The electric PF increases (decreases) with compressive (tensile) strain. The exception occurs in the case of $\lambda_0 = 10$ $\mu$m, in which the PFs reveal a non-monotonic behavior with strain $\epsilon_y$. It is worth mentioning that, for $\epsilon_{y} = 20\%$, the bottom of the conduction band of phosphorene surpasses the value of $0.7$ eV, and the Fermi energy used in Fig.~\ref{FigPF-Ele-Sy} becomes located inside the energy bandgap. In such a situation, the intraband term of the optical conductivity disappears, thereby surviving only the interband contribution, which produces abrupt reductions in the PFs. Finally, note that all SE rates tend to the free-space value at large distances, and the associated PFs are barely affected by strain, as expected.

\begin{figure}[t!]
	\centering
	 \includegraphics[width=0.75\linewidth]{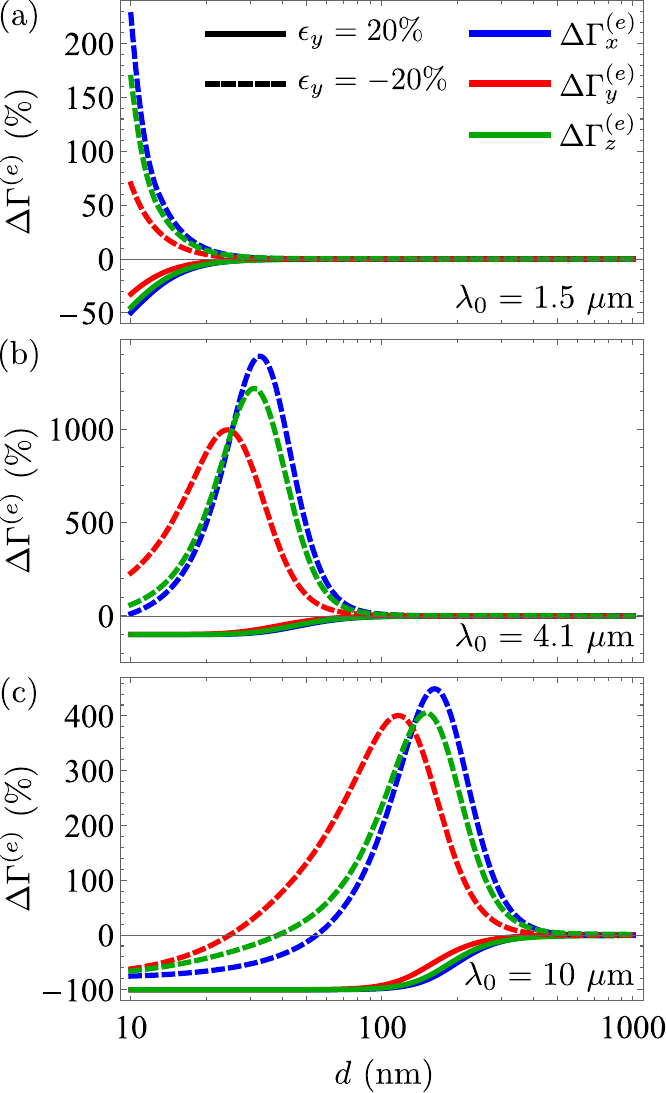}    
	\caption{Percentage variation in the electric PFs generated by the uniaxial strain along the $y$ direction as a function of the distance between the emitter and the phosphorene/SiC medium. Solid (dashed) curves show results for $\epsilon_y = 20\% (-20\%)$.}
	\label{FigDeltaPF-Ele-Sy}
\end{figure}

To quantify the degree of control of the SE, we define
\begin{equation}
	\Delta \Gamma^{(e)}_{\nu} = \frac{\Gamma^{(e)}_{\nu} \big|_{\epsilon_{x,y}\neq 0} - \Gamma^{(e)}_{\nu} \big|_{\epsilon_{x,y}=0}}{\Gamma^{(e)}_{\nu} \big|_{\epsilon_{x,y}=0}},
\label{deltaGamma}
\end{equation} 

\noindent where $\Gamma^{(e)}_{\nu}\big|_{\epsilon_{x,y}\neq 0}$ ($\Gamma^{(e)}_{\nu}\big|_{\epsilon_{x,y}=0}$) is the decay rate of the emitter aligned parallel to the $\nu$ direction near strained (relaxed) phosphorene/SiC half space. The percentage variation in the SE rates of the three emitters induced by strain applied in the $y$ direction for $\epsilon_{y}=\pm 20 \%$ as a function of separation between the emitter and the phosphorene/SiC half-space is illustrated in Fig.~\ref{FigDeltaPF-Ele-Sy}. From these results, the signature of the anisotropic nature of phosphorene becomes evident since $\Delta \Gamma^{(e)}_{x} \neq \Delta \Gamma^{(e)}_{y}$. We highlight that the electric PFs for $\lambda_0 = 4.1$ $\mu$m can be enhanced up to $1300 \%$ by compressive strain $\epsilon_y = -20 \%$. In the case of tensile strain $\epsilon_y = 20\%$, for which Fermi energy $E_{\rm F}=0.7$ eV lies inside the insulating gap, the PFs are reduced by a striking factor close to $100\%$, being nearly suppressed. In this situation, phosphorene becomes invisible to the emitter, demonstrating that strain can switch on and off quantum emission on demand. A residual Purcell effect still occurs due to the presence of the SiC substrate.

Despite the inherent anisotropic character of phosphorene, the effects of uniaxial strain along the $x$ direction are qualitatively similar when compared to the previous ones. By using an expression equivalent to Eq. (\ref{deltaGamma}), we can estimate the relative modification in the SE generated by strain applied in the $x$ direction, as presented in Appendix \ref{AppE}.


\subsection{Magnetic dipole emission \label{sec3}}

\begin{figure*}[t!]
	\centering
	\includegraphics[width=0.9\linewidth]{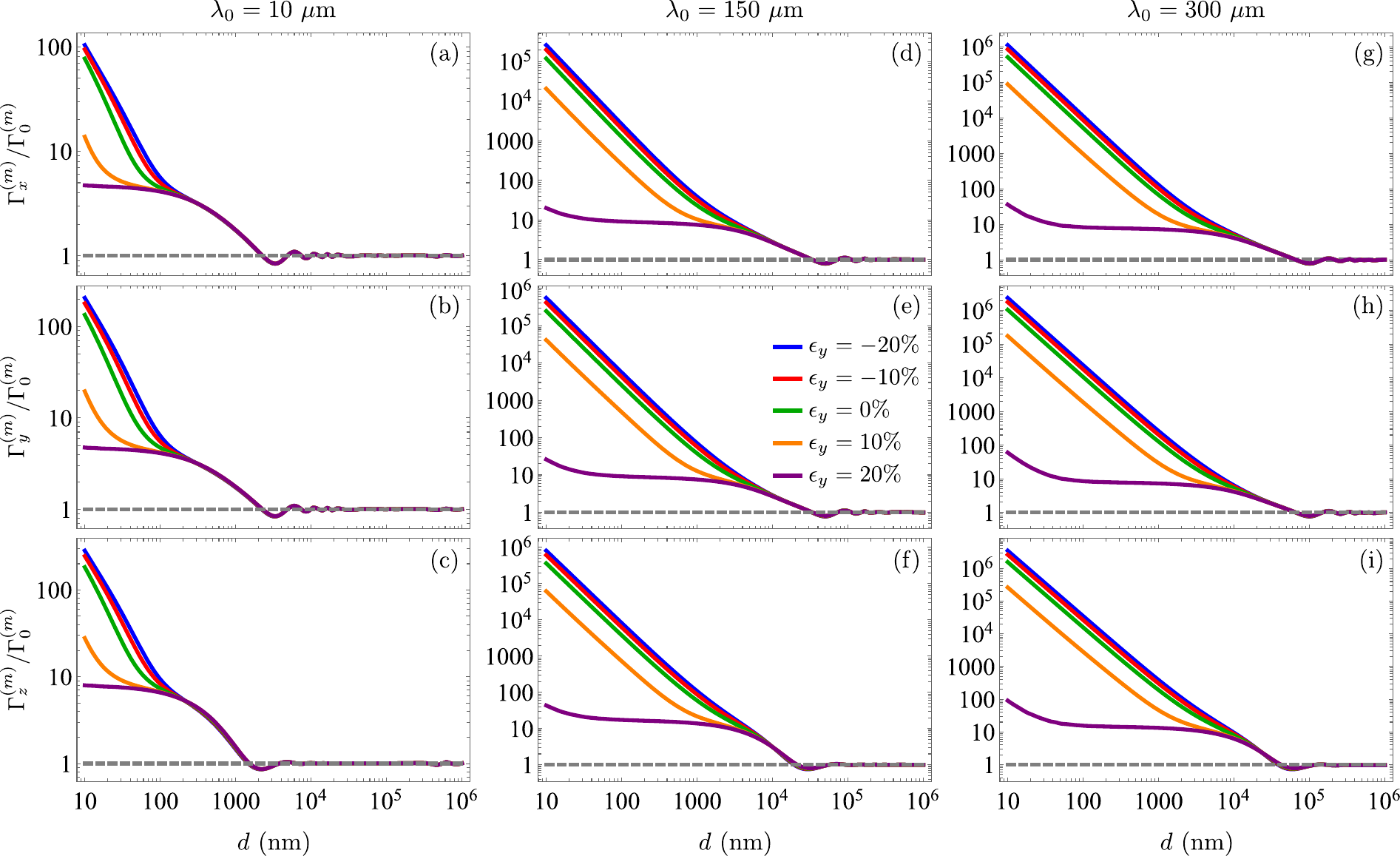}    
	\caption{Magnetic dipole Purcell factors as functions of the distance between the emitter and the phosphorene/SiC substrate medium. We considered the uniaxial strain in the $y$ direction with intensities $\epsilon_y=-20, -10, 0, 10, 20 \%$. The emitter's transition wavelengths are (a)-(c) $10$ $\mu$m, (d)-(f) $150$ $\mu$m, and (g)-(i) $300$ $\mu$m.}
	\label{FigPF-Mag-Sy}
\end{figure*}

We now discuss the study of the magnetic Purcell effect. The setup is similar to the one considered in Fig.~\ref{Fig-Boneco}. The difference is that the emitter decays to the ground state mediated by an MD transition. The magnetic PF can be obtained from \cite{Szilard_SE_VO2}
\begin{equation}
	\frac{\Gamma^{(m)} (\bm{r})}{\Gamma_0^{(m)}} = \frac{6 \pi c^3}{\omega_0^3} \,{\rm Im} \left[ \hat{m} \cdot \mathds{G}^{(m)} (\bm{r}, \bm{r}, \omega_0) \cdot \hat{m} \right] .
\label{MagneticPF}
\end{equation}

\noindent In the previous relation, $\Gamma_0^{(m)} = \mu_0 \omega_0^3 |\bm{m}|^2/3 \pi \hbar c^3$ is the free space SE rate of an MD emitter, $\bm{m}$ is the emitter's transition MD moment, $\hat{m} = \bm{m}/|\bm{m}|$, and $ \mathds{G}^{(m)} (\bm{r}, \bm{r}', \omega_0)$ is the magnetic Green dyadic. Analogously to the electric case, one can also express the magnetic PFs in terms of the diagonal part of the reflection matrices, and the formulas corresponding to the MD moments parallel to the $x$, $y$, and $z$ directions are 
\begin{align}
	\frac{\Gamma^{(m)}_{x}}{\Gamma_0^{(m)}} &= 1 + \frac{3}{4\pi k_0} \, {\rm Im} \left[ i \int d^2 \bm{k}_{\parallel} \frac{e^{2 i \sqrt{k_0^2 - k^2_{\parallel}} d}}{ k^2_{\parallel} \sqrt{k_0^2 - k^2_{\parallel}}}  \right. \nonumber \\ &\times \left. \left( k^2_y r_{pp} - \frac{k^2_x (k_0^2 - k^2_{\parallel})}{k^2_0} r_{ss} \right) \right],
\label{gammaMagXX} \\
	\frac{\Gamma^{(m)}_{y}}{\Gamma_0^{(m)}} &= 1 + \frac{3}{4\pi k_0} \, {\rm Im} \left[ i \int d^2 \bm{k}_{\parallel} \frac{e^{2 i \sqrt{k_0^2 - k^2_{\parallel}} d}}{k^2_{\parallel} \sqrt{k_0^2 - k^2_{\parallel}}} \right. \nonumber  \\ &\times \left. \left( k^2_x r_{pp} - \frac{k^2_y (k_0^2 - k^2_{\parallel})}{k^2_0} r_{ss} \right) \right],
\label{gammaMagYY} \\
	\frac{\Gamma^{(m)}_{z}}{\Gamma_0^{(m)}} &= 1 +  \frac{3}{4\pi k_0^3} \, {\rm Im} \left[ i \int d^2 \bm{k}_{\parallel} \frac{k^2_{\parallel} \, e^{2 i \sqrt{k_0^2 - k^2_{\parallel}} d}}{\sqrt{k_0^2 - k^2_{\parallel}}} r_{ss} \right]. 
\label{gammaMagZZ}
\end{align}

\noindent Note that the final expressions for the magnetic PFs are very similar to the electric ones, given in Eqs. (\ref{gammaElecXX})-(\ref{gammaElecZZ}), only requiring the exchange $r_{ss} \leftrightarrow r_{pp}$ \cite{Alaeian-Dionne-PRB-2015}. Likewise, $\Gamma^{(m)}_{x} \neq \Gamma^{(m)}_{y}$ due to the anisotropy of phosphorene.

In Fig.~\ref{FigPF-Mag-Sy}, we display the magnetic PFs as functions of the distance between the emitter and the phosphorene/SiC half-space for different values of uniaxial strain $\epsilon_y=-20, -10, 0, 10, 20 \%$ applied along the $y$ direction. We assume emitters with magnetic transition wavelengths $\lambda_0 = 10, 150, 300$ $\mu$m. The general behavior of the magnetic PFs presents some similarities when compared to the electric one, showing huge variations for small $d$. For larger $d$, the spontaneous decay rates tend to the free-space value, as expected. Furthermore, compressive strains ($\epsilon_y<0$) enhance the magnetic PFs, whereas tensile strains ($\epsilon_y>0$) diminish them. In this case, however, the magnetic PFs obey the scaling law $\Gamma^{(m)}_{\nu}/\Gamma^{(m)}_{0} \propto d^{-2}$ ($\nu = x, y, z$) for small separations, which can be clearly noticed in the plots with larger wavelengths ($\lambda_0=150$ $\mu$m and $\lambda_0=300$ $\mu$m) and for strain values whose Fermi energy $E_{\rm F}=0.7$ eV crosses the phosphorene bands ($\epsilon_{x,y}=-20, -10, 0, 10\%$). It is noteworthy that, for our choices of ED transitions, we did not find any scaling law in this same distance regime. We briefly mention that, in the case of ED emitters near graphene, it was shown that larger wavelength values and small distance regimes also obey a scaling law of the form $\Gamma^{(e)}/\Gamma^{(e)}_{0} \propto d^{-4}$ \cite{Gaudreau-Koppens-NanoLett-2013, Kort-Kamp-Amorim-FresnelCoefficients}.

\begin{figure}[t!]
    \centering
    \includegraphics[width=0.75\linewidth]{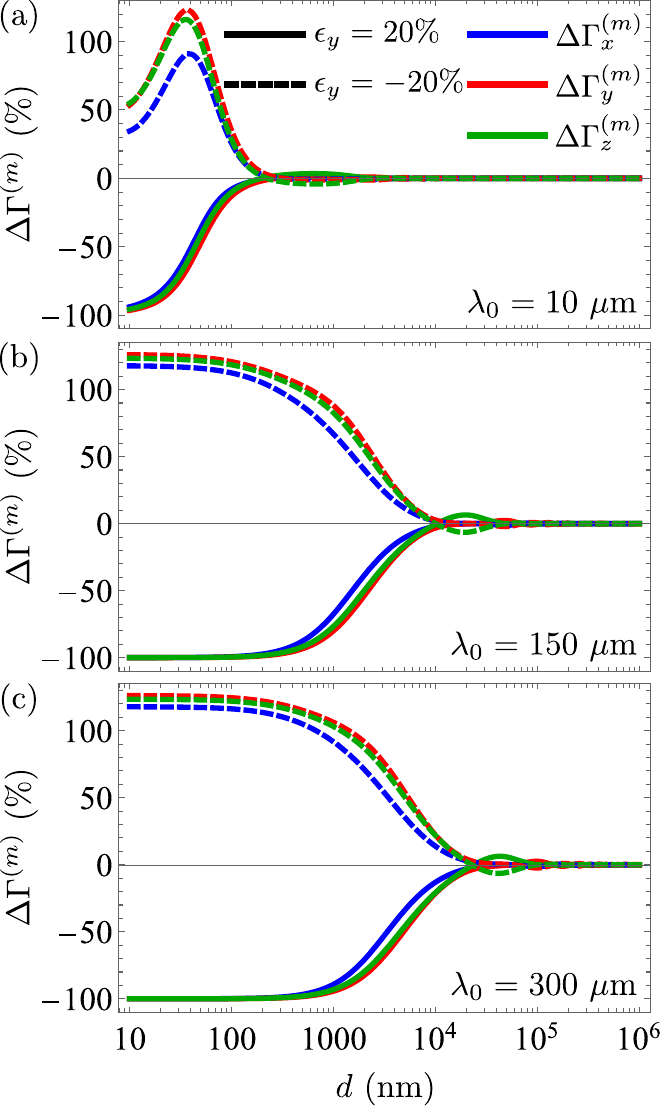}    
    \caption{Percentage variation in magnetic PFs generated by the uniaxial strain along the $y$ direction as a function of the distance between the emitter and the phosphorene/SiC substrate medium. Solid (dashed) curves show results for $\epsilon_y = 20\% (-20\%)$.}
    \label{FigDeltaPF-Mag-Sy}
\end{figure}

To quantify the change in the magnetic PFs produced by  strain, we define the quantity $\Delta \Gamma^{(m)}_{\nu}$ analogous to Eq. (\ref{deltaGamma}). Figure~\ref{FigDeltaPF-Mag-Sy} shows the results for the relative modification on the magnetic PFs produced by compressive (tensile) strain $\epsilon_y=-20\% (20\%)$. In Appendix \ref{AppE}, we included analogous plots considering strain along the $x$ direction. In both situations, the tensile strain may nearly suppress the magnetic PFs for small separations between the emitter and the phosphorene/SiC medium. The compressive strain along the two directions strongly enhances the magnetic PFs for small distances $d$ for the three wavelengths considered.


\section{Decay channels \label{sec4}} 

Results portrayed in Figs.~\ref{FigPF-Ele-Sy}-\ref{FigDeltaPF-Mag-Sy} demonstrate the potential of manipulating the electric and magnetic PFs of an emitter close to phosphorene/SiC by applying strain. To acquire more physical insights into these results, we analyze the decay channels of the emitted quanta in the specific case of dipoles perpendicular to the phosphorene interface with strain applied in the $y$ direction. The outcome is qualitatively alike when considering dipoles parallel to the surface and/or strain in the $x$ direction.

The relaxation process of an emitter in free space is followed by a radiative emission into propagating (Prop) modes detectable in the far field. When close to a given environment, other channels become accessible, especially in the near-field regime \cite{Kort-Kamp-Amorim-FresnelCoefficients, Szilard-PRB-2016}. For instance, the photon can be emitted into total internal reflection (TIR) modes that show up for $k_0 < k_{\parallel} < n_s k_0$, where $n_s = {\rm Re} \left[\sqrt{\varepsilon_s/\varepsilon_0} \right]$ stands for the medium refraction index. When losses are negligible, such modes propagate within the substrate but are evanescent in vacuum. Another possibility is the emitter to deexcite by a nonradiative process in which its energy is transferred directly to the half-space giving origin to lossy surface waves (LSWs). They emerge when $k_{\parallel} \gg n_s k_0$, their energy being quickly damped and converted into heat. From Eq. (\ref{gammaElecZZ}), we can extract the contributions of each channel to the decay rate as \cite{Kort-Kamp-Amorim-FresnelCoefficients, Szilard-PRB-2016}
\begin{align}
	\frac{\Gamma^{(e)}_{z,{\rm Prop}}}{\Gamma_0^{(e)}} &\simeq 1 + \frac{3}{4 \pi k^3_0} \int_0^{k_0} \!\! d k_{\parallel} \int_0^{2\pi} \!\! d\phi \frac{k^3_{\parallel} \, {\rm Re} \left[ e^{2 i \sqrt{k_0^2 - k^2_{\parallel}}d} r_{pp} \right]}{ \sqrt{k_0^2 - k^2_{\parallel}}} ,\label{gammaElecZZprop} \\
	\frac{\Gamma^{(e)}_{z,{\rm TIR}}}{\Gamma_0^{(e)}} &\simeq \frac{3}{4\pi k^3_0} \int_{k_0}^{n_s k_0} \!\! dk_{\parallel} \int_0^{2\pi} \!\! d\phi \frac{k^3_{\parallel} e^{-2 \sqrt{k^2_{\parallel} - k_0^2} d} \, {\rm Im} \left[r_{pp}\right]}{\sqrt{k^2_{\parallel} - k_0^2}},
\label{gammaElecZZtir} \\
	\frac{\Gamma^{(e)}_{z,{\rm LSW}}}{\Gamma_0^{(e)}} &\simeq \frac{3}{4\pi k^3_0} \int_{n_s k_0}^{\infty} \!\! d k_{\parallel} \int_0^{2\pi} \!\! d\phi \frac{k^3_{\parallel} e^{- 2 \sqrt{k^2_{\parallel} - k_0^2} d} \, {\rm Im} \left[r_{pp}\right]}{\sqrt{k^2_{\parallel} - k_0^2}}. 
\label{gammaElecZZlsw}
\end{align}

\noindent In the case of the magnetic Purcell effect, the decay contributions follow the aforementioned expressions with the exchange $r_{pp} \leftrightarrow r_{ss}$ [see Eq. (\ref{gammaMagZZ})]. The probabilities $p_{z,{\rm Prop}}^{(e)}$, $p_{z,{\rm TIR}}^{(e)}$, and $p_{z,{\rm Eva}}^{(e)}$ of energy emission in the different decay channels are calculated by the ratio between the partial and the total rates. Similar decomposition can be done for dipoles lying parallel to the $x$ and $y$ directions.

\begin{figure}[t!]
	\centering
	\includegraphics[width=0.75\linewidth]{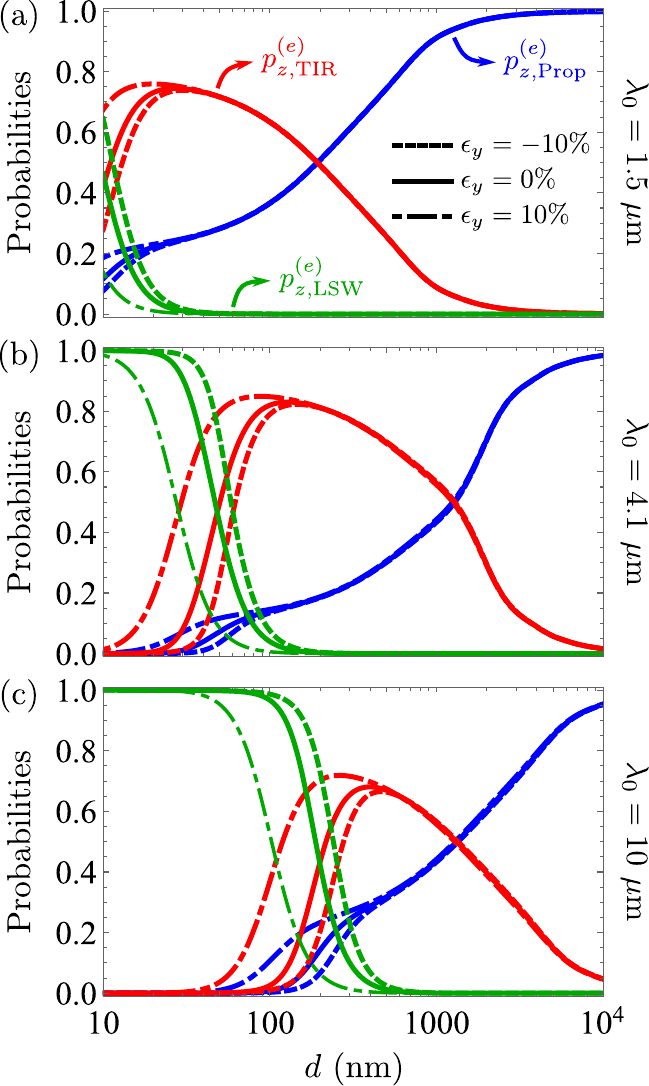}    
	\caption{Decay channels probabilities of an ED as a function of the distance $d$ for strain $\epsilon_y = -10, 0, 10\%$.}
	\label{Decay-Chanels-ED_zz}
\end{figure}

\begin{figure}[t!]
	\centering
	\includegraphics[width=0.75\linewidth]{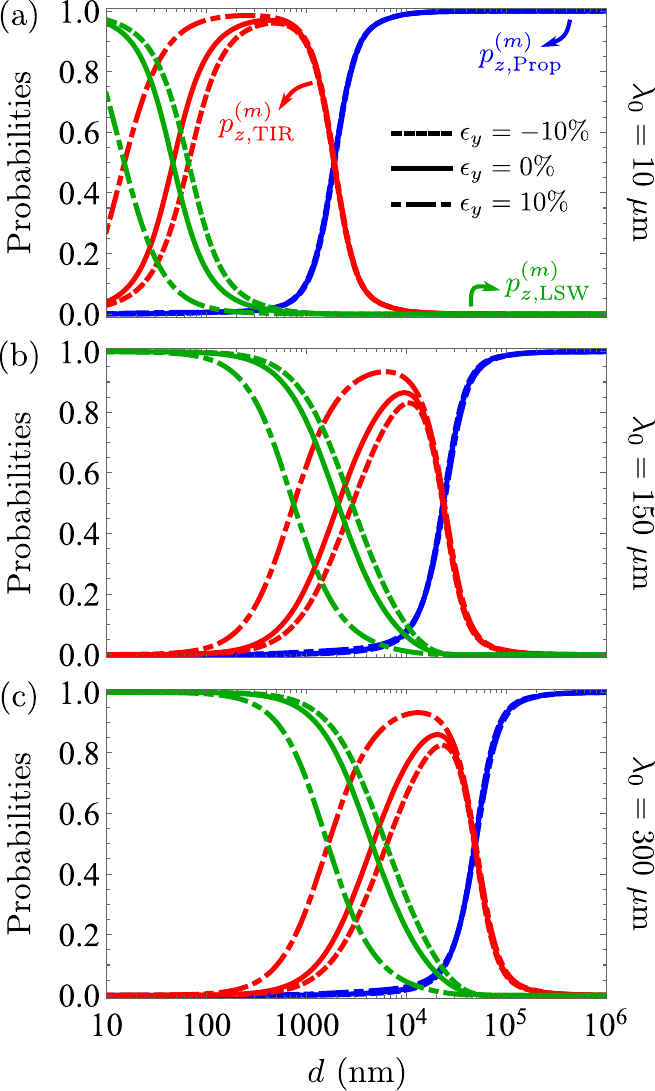}    
	\caption{Decay channels probabilities of an MD as a function of the distance $d$ for strain $\epsilon_y = -10, 0, 10\%$.}
	\label{Decay-Chanels-MD_zz}
\end{figure}

In Fig.~\ref{Decay-Chanels-ED_zz}, we depict the decay probabilities as functions of the distance $d$ in order to uncover the role of the different relaxation channels for an ED emitter. Each plot refers to a transition wavelength ($\lambda_0 = 1.5, 4.1$, and $10$ $\mu$m), and different strain intensities along the $y$ direction ($\epsilon_y=-10, 0, 10\%$) are shown in each panel. As $d$ increases, the propagating modes become the dominant decay channel, minimizing the effects of the interface on SE. This can be clearly noticed for $\lambda_0 = 1.5$ $\mu$m, in which case the decay via propagating modes dominates. However, the same behavior will also occur for the other wavelengths provided $d$ is large enough. Indeed, as $d$ decreases the propagating channel gets progressively suppressed, giving rise to competition between TIR and LSW modes. Moreover, the probabilities associated with these decay channels may be highly influenced by strain to the point where one may tune the relative dominance between TIR and LSW processes. For the transition wavelength $\lambda_0 = 4.1$ $\mu$m, this variation in the dominant decay channel can be achieved for separations $20$ nm $\lesssim d \lesssim 100$ nm, while for $\lambda_0 = 10$ $\mu$m, the corresponding range is $100$ nm $\lesssim d \lesssim 300$ nm. Lastly, note that LSWs govern the SE in the near-field regime (which also holds for $\lambda_0 = 1.5$ $\mu$m in the extreme near-field). In Fig.~\ref{Decay-Chanels-MD_zz}, we display the different relaxation channels probabilities for the MD case for the transition wavelengths $\lambda_0 = 10, 150, 300$ $\mu$m. The main aspects of the discussion follow analogously to the previous case, with the difference that the distance scales for which each mode is most relevant may comprise larger values. Ultimately, Figs.~\ref{Decay-Chanels-ED_zz} and \ref{Decay-Chanels-MD_zz} unveil the possibility of controlling the preferable pathway of emitted energy in the decay process via uniform uniaxial strain. It also shows that, at a fixed distance, emitters with larger wavelengths are more prone to the control of spontaneous emission by strain in phosphorene. 

\section{Conclusions \label{sec5}}

In summary, we have applied a tight-binding approach that goes beyond the low-energy description traditionally used in nanophotonics to investigate spontaneous emission in phosphorene layers. With this methodology, we demonstrate remarkable external control over the electric and magnetic Purcell effects by applying uniform strain. The application of strain is also shown to control the different decay pathways that contribute to SE. The use of high-strain levels is only possible due to the great flexibility of the phosphorene sheet that has its origins in its puckered lattice structure. The strain-based approach to control quantum emission in phosphorene is within the reach of state-of-the-art techniques \cite{Exp-Strain-Ph-1, Exp-Strain-Ph-2}, and it is a clear advantage when compared to existing proposals based on electromagnetic fields acting as external agents. We hope that our results will not only allow for an alternative method to tune spontaneous emission but also be relevant in developing new photonic devices, as the Purcell effect is a key mechanism in many quantum-optical applications such as single-photon sources.

\section*{Acknowledgments}
T.P.C., F.A.P., F.S.S.R., and C.F. thank the Brazilian Agencies CAPES, CNPq, and FAPERJ for financial support. P.P.A. is supported by the São Paulo Research Foundation (FAPESP) through Grant No. 2021/04861-7. W.J.M.K.-K. acknowledges the Laboratory Directed Research and Development program of Los Alamos National Laboratory under Projects No. 20220228ER and 20220627DI. T.P.C. would like to thank R. de Melo e Souza for the fruitful discussions.


\appendix
\section{Tight-binding model of phosphorene \label{AppA}}

Throughout this work, we describe the electronic structure of phosphorene by employing a simplified two-band tight-binding model \cite{Rudenko-Katsnelson-Ph-NoStrain, Rodin-Carvalho-CastroNeto-Ph-NoStrain}. The inclusion of uniform strain is done by using the Harrison prescription \cite{Peeters-StrainPhosphorene-Model, Midtvedt-Lewenkopf-Croy-JPCM, HarrisonBook}. In short, this model captures the behavior of the anisotropic spectra of phosphorene with a uniform strain field. The Hamiltonian can be cast into
\begin{eqnarray}
	H^{(2)}_{\bm{q}}=\begin{bmatrix}
B_{\bm{q}} e^{i(q_a-q_b)/2} & A_{\bm{q}} + C_{\bm{q}} e^{i(q_a-q_b)/2}  \\
A^*_{\bm{q}} + C^*_{\bm{q}} e^{-i(q_a-q_b)/2} & B_{\bm{q}} e^{i(q_a-q_b)/2} 
\end{bmatrix}, \nonumber \\
\label{HP2}
\end{eqnarray}

\noindent where
\begin{align}
A_{\bm{q}} &= t_2+t_5 e^{-iq_a}, \\
B_{\bm{q}} &= 4t_4 e^{-i(q_a-q_b)/2}\cos \left(\frac{q_a}{2}\right) \cos \left(\frac{q_b}{2}\right), \\
C_{\bm{q}} &= 2 e^{iq_b/2}\cos \left(\frac{q_b}{2}\right)(t_1 e^{-iq_a}+t_3).
\end{align}

\noindent Here, $q_a={\bm{q}} \cdot {\bm{a}}$, $q_b={\bm{q}} \cdot {\bm{b}}$, where ${\bm{a}}=\left(4.580 \angstrom\right) \hat{x}$ and ${\bm{b}}=\left(3.320 \angstrom\right) \hat{y}$ are lattice vectors of the unstrained phosphorene monolayer and ${\bm{q}}$ is the electronic momentum. One can follow the Harrison prescription and include the effect of strain in the hopping amplitudes \cite{HarrisonBook}
\begin{equation}
t_{i}\approx \left( 1-2 \alpha^i_x\epsilon_x- 2\alpha^i_y \epsilon_y-2\alpha^i_z \epsilon_z \right)t^0_i, \label{tHarrison}
\end{equation}

\noindent with $t^0_1=-1.220$ eV,  $t^0_2=3.665$ eV, $t^0_3=-0.205$ eV, $t^0_4=-0.105$ eV, and $t^0_5=-0.055$ eV being hopping parameters of the unstrained phosphorene \cite{Rudenko-Katsnelson-Ph-NoStrain, Rodin-Carvalho-CastroNeto-Ph-NoStrain}, and $\alpha^i_{\mu}=\left(\delta^i_{\mu}/|\bm{\delta}^i|\right)^2$, where $\bm{\delta}^i$ is the $i$-th hopping vectors: $\bm{\delta}^1=\left(r^0_{1x}, r^0_{1y}, 0\right)$, $\bm{\delta}^2=\left(-r^0_{2x}, 0, -r^0_{2z}\right)$, $\bm{\delta}^3=\left(-2r^0_{2x}-r^0_{1x}, r^0_{1y}, 0\right)$, $\bm{\delta}^4=\left(r^0_{1x}+r^0_{2x}, r^0_{1y}, -r^0_{2z}\right)$, and $\bm{\delta}^5=\left(2r^0_{1x}+r^0_{2x}, 0, -r^0_{2z}\right)$. They are written in terms of vectors $\bm{r}^0_1=\left( 1.503, \ 1.660, \ 0\right) \angstrom$ and $\bm{r}^0_2=\left( 0.786, \ 0, \ 2.140\right) \angstrom$. The parameter $\epsilon_{\mu}$ is negative (positive) for compressive (tensile) uniaxial strain along the $\mu$ direction ($\mu=x, y, z$).

\begin{figure}[t!]
	\centering
	 \includegraphics[width=0.9\linewidth]{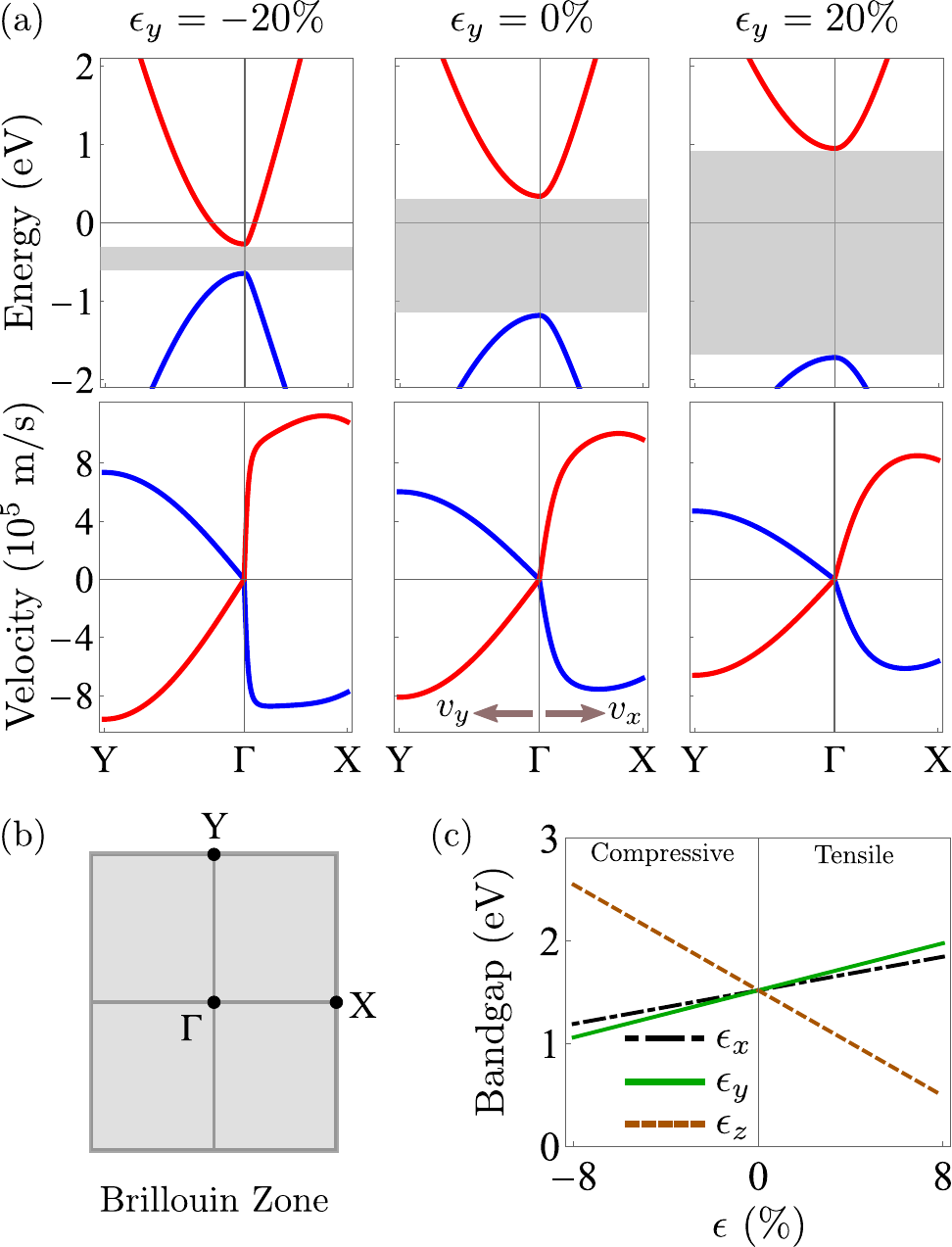}    
	\caption{(a) The three panels represent the energy (top) and velocities (bottom) of the carriers computed for different values of uniaxial strain applied in the $y$ direction $\epsilon_y= -20 \ \text{(compressive)}, 0 \ \text{(unstrained)}, 20 \ \text{(tensile)} \%$ obtained from the two-bands model. The shaded region represents the bandgap in energy spectra. The plots were done in the path $\bm{q} \in [Y: (0,\pi/|\bm{b}|) \rightarrow \Gamma: (0, 0) \rightarrow X: (\pi/|\bm{a}|,0)]$ of the Brillouin zone, shown in panel (b). In the path $\text{Y} \rightarrow \Gamma$ ($\Gamma \rightarrow  X$), we plotted the component $v_y$ ($v_x$) of the electronic velocity. (c) The bandgap in energy spectra as a function of uniaxial strain applied in three directions.}
	\label{FigSpectra}
\end{figure}

In Fig.~\ref{FigSpectra}(a), we show how strain along the $y$ direction modifies the energy spectra $E(\bm{q})$ and the velocity of the carriers $\bm{v}(\bm{q})=\hbar^{-1}\nabla_{\bm{q}} E(\bm{q})$. The compressive strain ($\epsilon_y<0$) reduces the energy gap of phosphorene at the $\Gamma$ point and enhances the modulus of the velocity of the carriers. On the other hand, the tensile strain ($\epsilon_y>0$) enhances the energy gap of the electronic spectra and reduces the velocity of the carriers. The behaviors of the energy spectra and the electronic velocity with strain along the $x$ direction are qualitatively similar, while strain in the $z$ direction produces an opposite effect, as can be seen in Fig. \ref{FigSpectra}(c).


\section{Optical Conductivity \label{AppB}}

With Hamiltonian (\ref{HP2}), we can compute the matrix elements of the optical conductivity tensor of strained phosphorene, written in Eq. (\ref{TensorSigma}). Generally, it is possible to express the optical conductivity as a sum of two contributions, to wit, $\sigma_{\mu,\mu}(\omega, \epsilon_{\mu})=\sigma^{\text{(Inter)}}_{\mu,\mu}(\omega)+\sigma^{\text{(Intra)}}_{\mu,\mu}(\omega)$  \cite{Novko-Opticond-Phosphorene,Moshayedi-Opticond-Phosphorene}. The intraband contribution is given by 
\begin{equation}
	\sigma^{\text{(Intra)}}_{\mu, \mu} (\omega) =  \frac{i D_{\mu,\mu}}{\hbar \omega + i \eta_1},
\label{SigmaIntra}
\end{equation}

\noindent where the Drude weight is
\begin{equation}
	D_{\mu, \mu} = - g_s \frac{e^2 \hbar}{S} \sum_{n = 1,2} \sum_{\bm{q}} f'_{n,\bm{q}} \bra{u_{\bm{q},n}} \hat{v}_{\mu} (\bm{q}) \ket{u_{\bm{q}, n}}^2. 
\label{DW}
\end{equation}

\noindent The interband contribution is obtained from the Kubo formula \cite{Low-Rodin-Optcond}
\begin{align}
	\sigma^{\text{(Inter)}}_{\mu, \mu} (\omega) &= i g_s \frac{e^2\hbar}{S} \sum_{\bm{q}} \frac{\bra{u_{\bm{q},1}} \hat{v}_{\mu} (\bm{q}) \ket{u_{\bm{q}, 2}}^2}{\Delta E_{\bm{q}}} \nonumber \\
	&\times \left[\frac{f_{\bm{q},1} - f_{\bm{q},2}}{\hbar  \omega + \Delta E_{\bm{q}} + i \eta_2} + \frac{f_{\bm{q},1} - f_{\bm{q},2}}{\hbar \omega - \Delta E_{\bm{q}} + i \eta_2} \right],
\label{SigmaInter}
\end{align} 

\begin{figure}[b!]
	\centering
	 \includegraphics[width=0.7\linewidth]{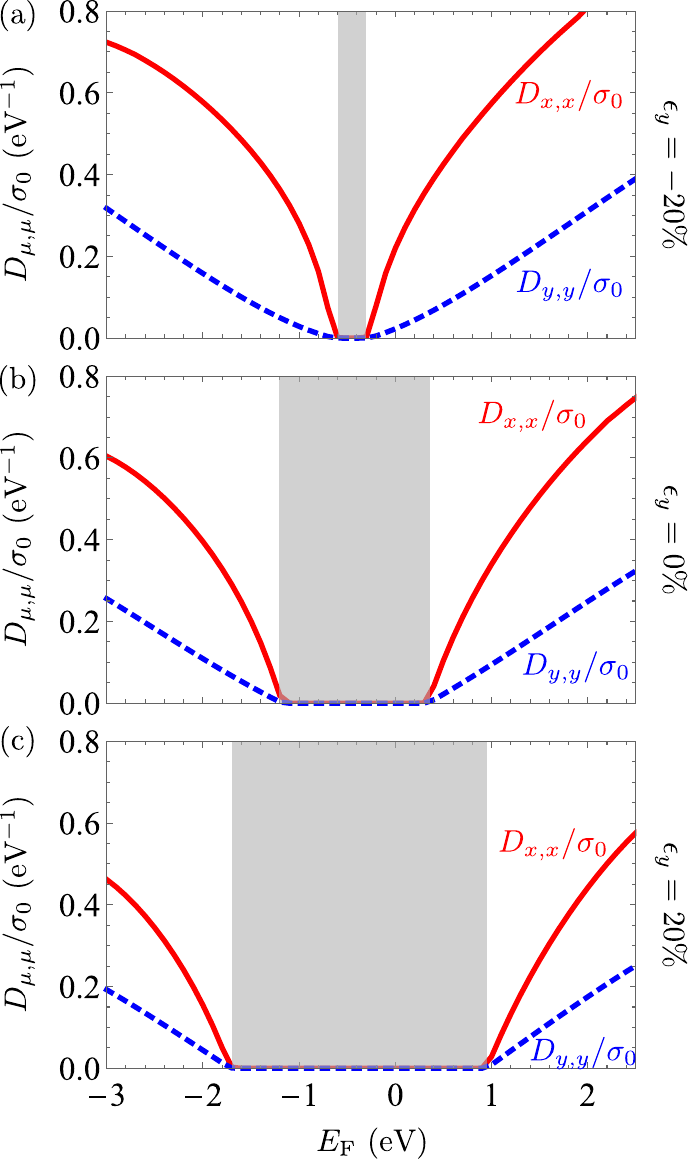}    
	\caption{Drude weights $D_{x,x}$ (solid red) and $D_{y,y}$ (dashed blue) as functions of the Fermi energy for distinct values of strain applied in the $y$ direction $\epsilon_y= -20$ (top), $0$ (center), $20\%$ (bottom). Again, the shaded region represents the bandgap in the energy spectra.}
	\label{FigDW}
\end{figure}

\noindent where $\ket{u_{\bm{q}, 1(2)}}$ is the eigenvector of the Hamiltonian (\ref{HP2}), associated with energy bands $E_{\bm{q}, 1(2)}$, and $\Delta E_{\bm{q}} = \left(E_{\bm{q}, 2}-E_{\bm{q}, 1}\right)$. Furthermore, $f_{\bm{q},1(2)}=f_{\text{FD}}(E_{\bm{q}, 1(2)})$, with $f_{\text{FD}}(E)=\big\{\exp[(E-E_{F})/k_\text{B}T] +1\big\}^{-1}$ being the Fermi-Dirac distribution. In Eq. (\ref{DW}), we also defined $f'_{1(2),\bm{q}}=\left[ \partial f_{\text{FD}}(E)/\partial E \right] \big|_{E=E_{\bm{q},1(2)}}$. The velocity operator in the $\mu$ direction is given by $\hat{v}_{\mu}(\bm{q})=\hbar^{-1}\partial H^{(2)}_{\bm{q}}/\partial q_{\mu}$, with $\mu=x, y$. In Eqs. (\ref{DW}) and (\ref{SigmaInter}), $g_s=2$ is the spin degeneracy factor, $S$ is the area of the phosphorene layer. We express the results in terms of $\sigma_0=e^2/\hbar$. In Eq. (\ref{SigmaIntra}), $\eta_1=\hbar/(2\tau)$ and $\tau$ is the momentum relaxation time \cite{Cysne-Rappoport-DisorderGraphene}. In Eq. (\ref{SigmaInter}), $\eta_2$ is a small phenomenological quantity. In all results presented in this paper, we used $T= 180$ K, $\eta_1= 25$ meV, and $\eta_2= 25$ meV \cite{Novko-Opticond-Phosphorene, Moshayedi-Opticond-Phosphorene, Zhu-Zhang-Li-Relaxation-Time, Lv-Lu-Relaxation-Time}.

Now, we briefly discuss the optical conductivity in insulating and metallic cases. In Fig.~\ref{FigDW}, we show the Drude weight as a function of the Fermi energy for different values of uniaxial strain along the $y$ direction. These plots illustrate how the Drude weight can be well controlled by uniform strain, which occurs as a direct consequence of the change in the velocities of the carriers due to the strain, as previously mentioned in Fig.~\ref{FigSpectra}(a). In addition, we may separate two distinct situations depending on the Fermi energy. In the insulating case, $E_{\rm F}$ lies inside the energy gap (shaded region in Fig.~\ref{FigDW}), and the Drude weight vanishes. Consequently, the intraband term does not contribute to the optical conductivity. The metallic case occurs when $E_{\text{F}}$ crosses a Bloch band of the phosphorene energy spectra. In this situation, the Drude weight is non-zero, and the optical conductivity has contributions from both interband and intraband terms.

\begin{figure}[b!]
	\centering
	\includegraphics[width=0.99\linewidth]{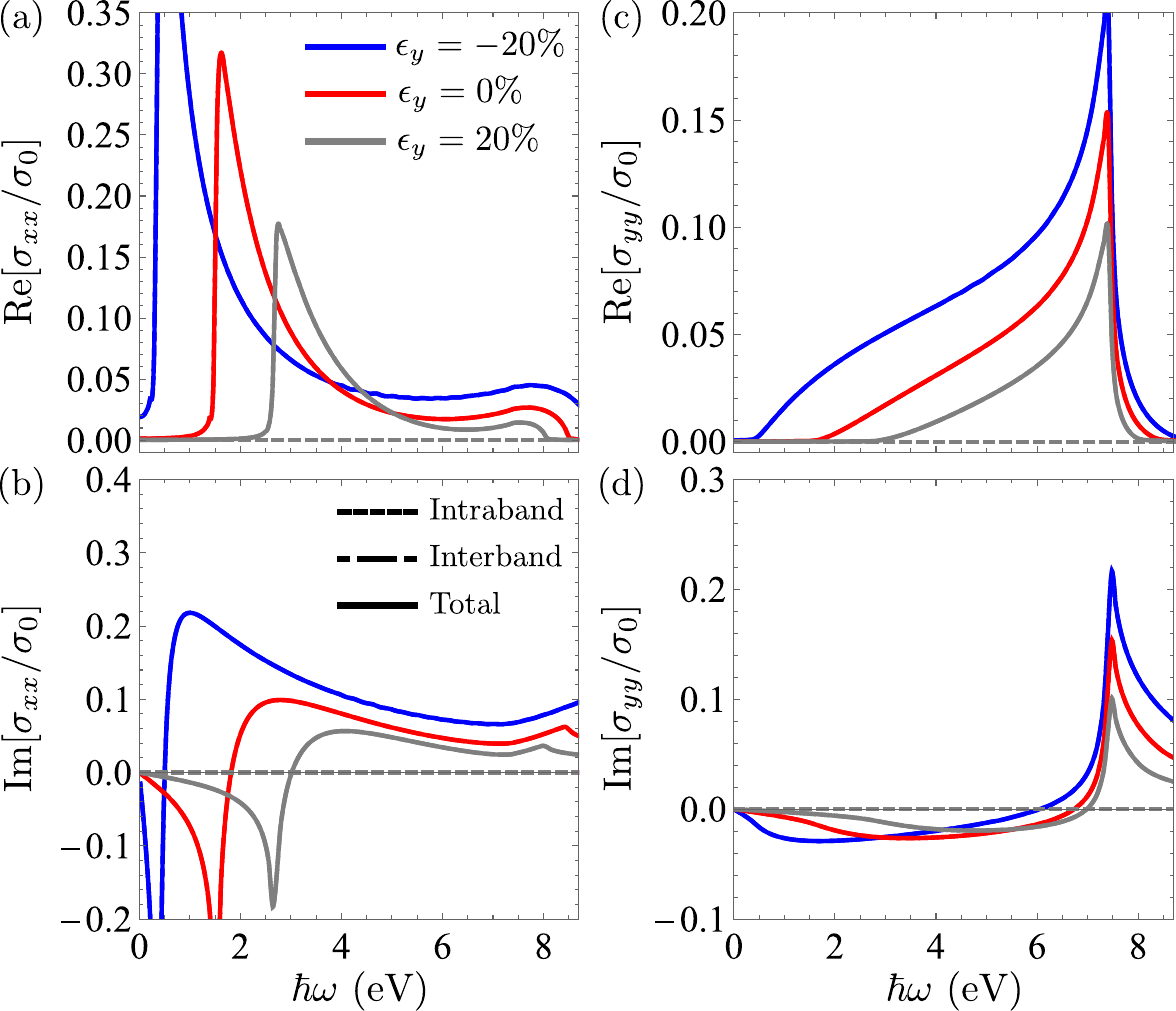}    
	\caption{Real and imaginary parts of phosphorene's optical conductivity in the insulating regime ($E_{\rm{F}}$ inside energy gap) under the effect of different uniaxial strains: $\epsilon_y=-20$ (blue), $0$ (red), and $20\%$ (gray). Dashed curves represent intraband contributions, circles represent interband contributions, and solid lines represent the total optical conductivity.}
	\label{FigSigma-insulating}
\end{figure}

\begin{figure}[t!]
	\centering
        \includegraphics[width=0.99\linewidth]{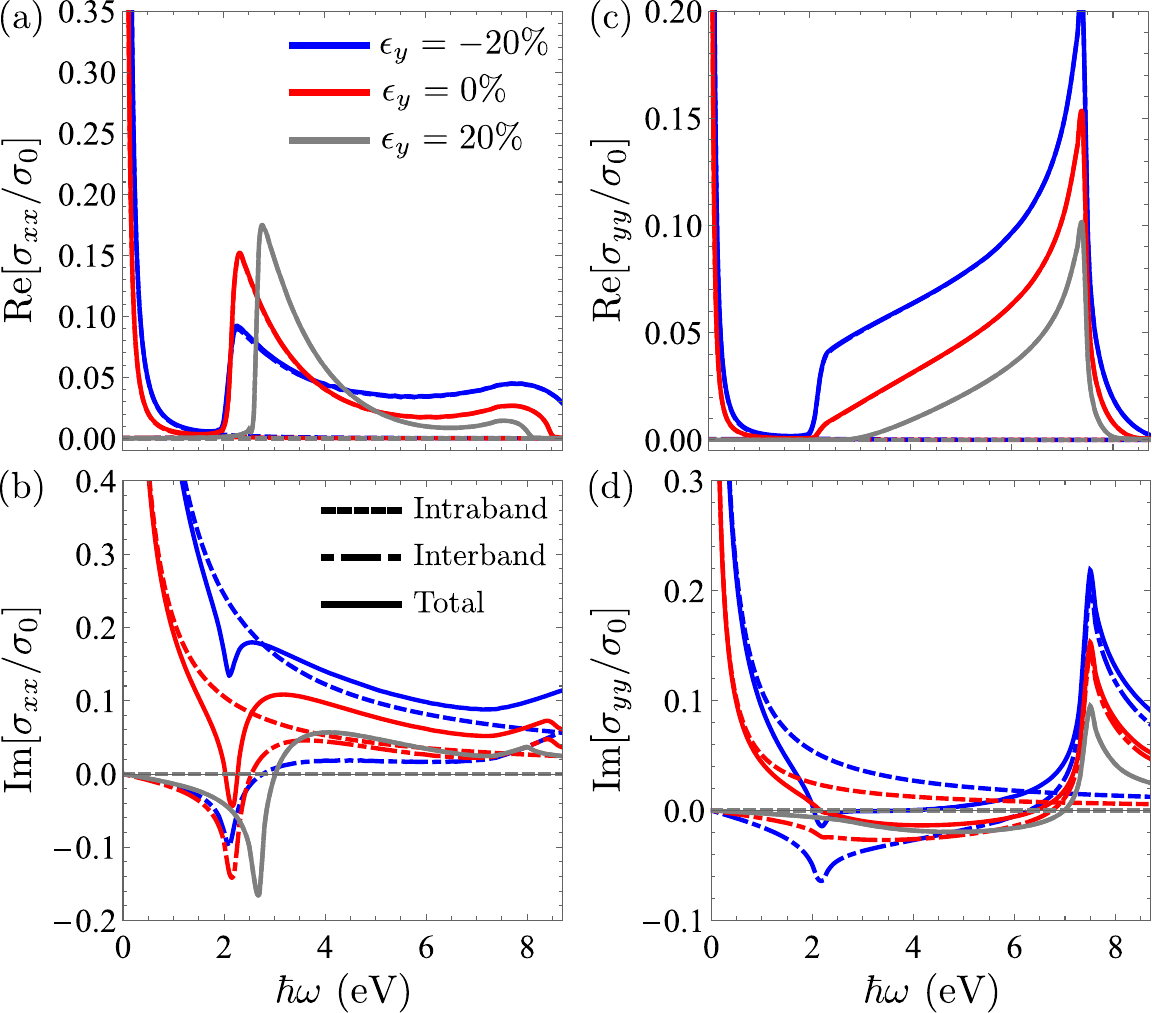}    
	\caption{Real and imaginary parts of phosphorene's optical conductivity for $E_{\rm{F}}= 0.7$ eV under the effect of different strains: $\epsilon_y=-20\%$ (blue) (metallic), $0\%$ (red) (metallic), and $20\%$ (gray) (insulating). Dashed curves represent intraband contributions, circles represent interband contributions, and solid lines represent the total optical conductivity.}
	\label{FigSigma-metallic}
\end{figure}

In Fig.~\ref{FigSigma-insulating}, we show the real and imaginary parts of the optical conductivity for the case of Fermi energy $E_{\rm F}= (E_{\bm{q}=0,2}+E_{\bm{q}=0,1})/2$ lying inside the insulating bandgap and different values of $\epsilon_y$. In this situation, $\sigma_{\mu,\mu} (\omega) = \sigma^{\text{(Inter)}}_{\mu,\mu} (\omega)$. For comparison, we show in Fig.~\ref{FigSigma-metallic} the same quantities, but for a fixed Fermi energy $E_{\rm F} = 0.7$ eV. For $\epsilon_y=-20\%$ and $\epsilon_y=0\%$ in Fig.~\ref{FigSigma-metallic}, the Fermi energy crosses the electronic bands of phosphorene, and the system exhibits a metallic behavior, such that $\sigma_{\mu,\mu}(\omega) = \sigma^{\text{(Inter)}}_{\mu,\mu}(\omega) + \sigma^{\text{(Intra)}}_{\mu,\mu}(\omega)$. For $\epsilon_y=20 \%$, the bottom of the conduction band surpasses the Fermi energy $E_{\rm F}=0.7$ eV that enters into the energy gap region, thereby vanishing the intraband contribution to the conductivity. From these results, it becomes evident the possibility of controlling the optical responses of phosphorene by means of uniaxial strain. Typically, fixed Fermi energy can be maintained by controlling the carriers doping \cite{Kort-Kamp-Amorim-FresnelCoefficients, Guinea-Martin-Moreno-PRR2019}, which is possible by tuning the back gate voltage in the substrate \cite{Das-Roelofs-ACS-Nano-2014}.


\section{Reflection Coefficients \label{AppC}}

In our system, the phosphorene sheet is grown on top of a substrate of silicon carbide (SiC), whose electrical permittivity can be modeled by a simple Drude-Lorentz model \cite{DL-SiC-PaliK_book}
\begin{equation}
	\frac{\varepsilon_{\text{SiC}}(\omega)}{\varepsilon_0} = \varepsilon_{\infty} \left( 1+ \frac{\omega^2_L-\omega^2_T}{\omega_T^2 - \omega^2 - i\omega/\tau_{\text{SiC}}} \right),
\end{equation} 

\noindent with $\varepsilon_{\infty}=6.7$, $\omega_{L}=182.7 \times 10^{12}$ rad/s, $\omega_T=149.5 \times 10^{12}$ rad/s, and $\tau_{SiC}^{-1}=0.9 \times 10^{12}$ rad/s.

The reflection coefficients of the phosphorene/SiC medium can be derived by solving the Maxwell equations with proper boundary conditions \cite{Moreno-FresnelCoefficients, Kort-Kamp-Amorim-FresnelCoefficients}. Following Ref. \cite{Kort-Kamp-Amorim-FresnelCoefficients}, we obtain the diagonal parts of the reflection matrices
\begin{equation}
	r_{pp} = \frac{\Delta_+^{\rm T} \Delta_-^{\rm L} + \Lambda^2}{\Delta_+^{\rm T} \Delta_+^{\rm L} + \Lambda^2} \;\;\;\; {\rm and} \;\;\;\; r_{ss} = - \frac{\Delta_-^{\rm T} \Delta_+^{\rm L} + \Lambda^2}{\Delta_+^{\rm T} \Delta_+^{\rm L} + \Lambda^2}.
\label{RssRpp}
\end{equation}  

\noindent In both equations,
\begin{align}
	\Delta^{\rm L}_{\pm} &= \left( k_{z,1} \varepsilon_2 \pm k_{z,2} \varepsilon_1 + k_{z,1} k_{z,2} \sigma_{\rm L}/\omega \right)/\varepsilon_0,
\label{EqDeltaL}\\
	\Delta^{\rm T}_{\pm} &= \left(k_{z,2} \mu_1 \pm k_{z,1} \mu_2 +\omega \mu_1 \mu_2 \sigma_{\rm T} \right)/\mu_0,
\label{EqDeltaT}\\
	\Lambda^2 &= - Z_0^2 \mu_1 \mu_2 k_{z,1} k_{z,2} \sigma_{\rm LT}^2/\mu_0^2.
\label{EqLambda}
\end{align}

\noindent In our system, medium $1$ is vacuum ($\varepsilon_1 = \varepsilon_0$, $\mu_1 = \mu_0$) and medium $2$ is the SiC substrate ($\varepsilon_2 = \varepsilon_{\text{SiC}}$, $\mu_2 = \mu_0$). In Eqs. (\ref{EqDeltaL})-(\ref{EqLambda}), $Z_0 = \sqrt{\mu_0/\varepsilon_0}$, $k_{z,n} = \sqrt{k^2_n - k^2_{\parallel}}$, $k_{\parallel} = |{\bm{k}_{\parallel}}| = |k_x \hat{x} + k_y \hat{y}|$, and $k_n = \omega \sqrt{\varepsilon_n \mu_n}$ ($n=1,2$). We have also defined the optical conductivities in the reference frame of the incident electromagnetic wave \cite{Kort-Kamp-Amorim-FresnelCoefficients}, so that $\sigma_{\rm L}=\left( k_x^2 \sigma_{xx} + k_y^2 \sigma_{yy} \right)/k^2_{\parallel}$, $\sigma_{\rm T} = \left( k_y^2 \sigma_{xx} + k_x^2 \sigma_{yy} \right)/k^2_{\parallel}$, and $\sigma_{\rm LT}=k_x k_y \left( \sigma_{yy} - \sigma_{xx} \right)/k_{\parallel}^2$, where $\sigma_{xx (yy)}$ are given by Eqs. (\ref{SigmaIntra})-(\ref{SigmaInter}). We stress that the inclusion of substrate in this work has conceptual importance, allowing for the application of strain in the plane of phosphorene. Nevertheless, the optical response in the phosphorene/SiC half-space is dominated by phosphorene. The strain along the $z$ direction cannot be controlled in the setup proposed in Fig.~\ref{Fig-Boneco}.


\begin{figure*}[t!]
	\centering
	\includegraphics[width=0.9\linewidth]{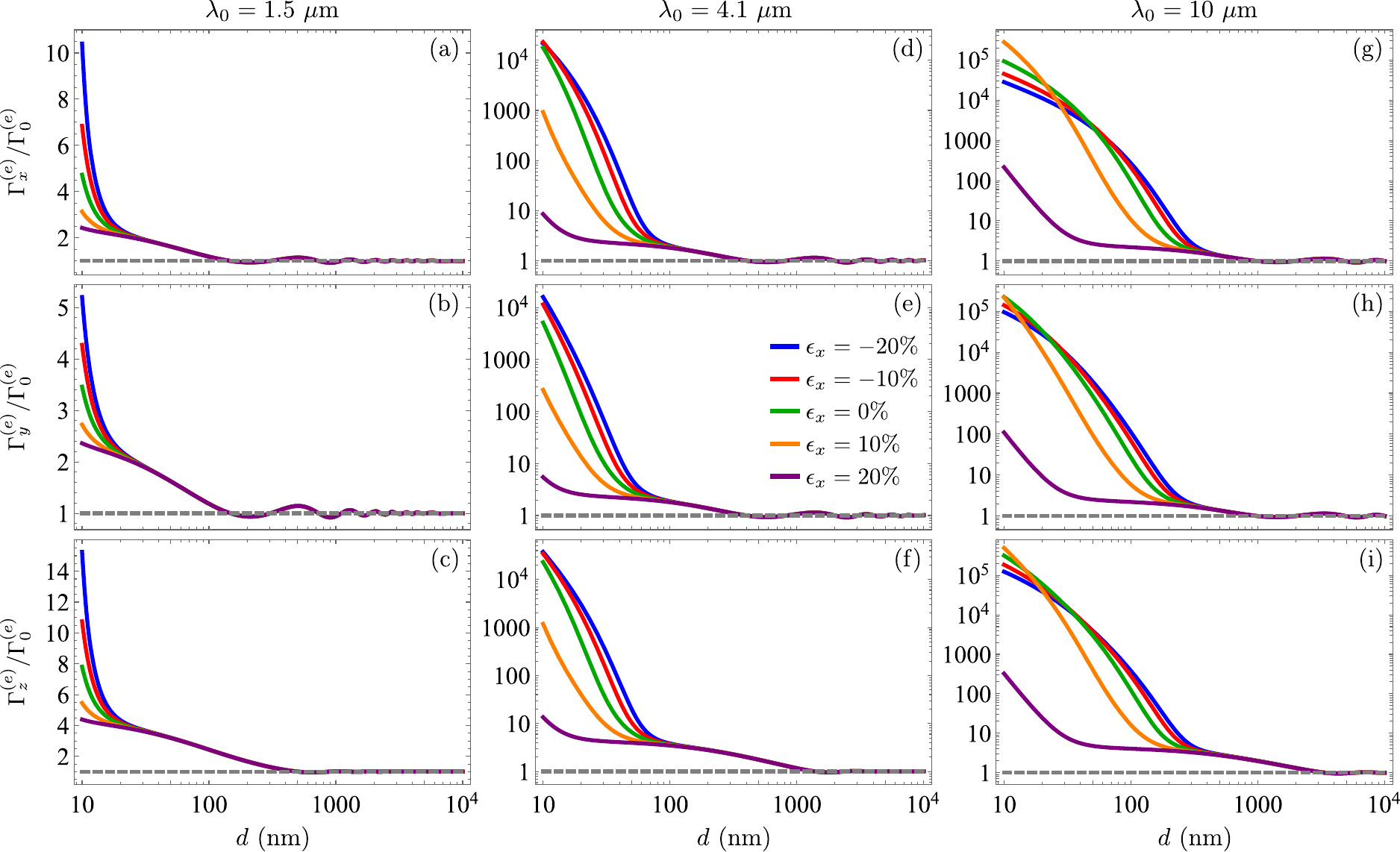}    
	\caption{Electric dipole Purcell factors $\Gamma^{(e)}_x/\Gamma_0^{(e)}$, $\Gamma^{(e)}_y/\Gamma_0^{(e)}$, and $\Gamma^{(e)}_z/\Gamma_0^{(e)}$ as functions of the distance $d$ between the emitter and the phosphorene/SiC medium. Here, we considered the uniaxial strain in the $x$ direction with intensities $\epsilon_x=-20, -10, 0, 10, 20 \%$. The emitter's transition wavelengths are (a)-(c) $1.5$ $\mu$m, (d)-(f) $4.1$ $\mu$m, and (g)-(i) $10$ $\mu$m.}
	\label{FigPF-Ele-Sx}
\end{figure*}


\section{Purcell factors for strains in the $x$ direction \label{AppE}}

\begin{figure*}[t!]
	\centering
	\includegraphics[width=0.9\linewidth]{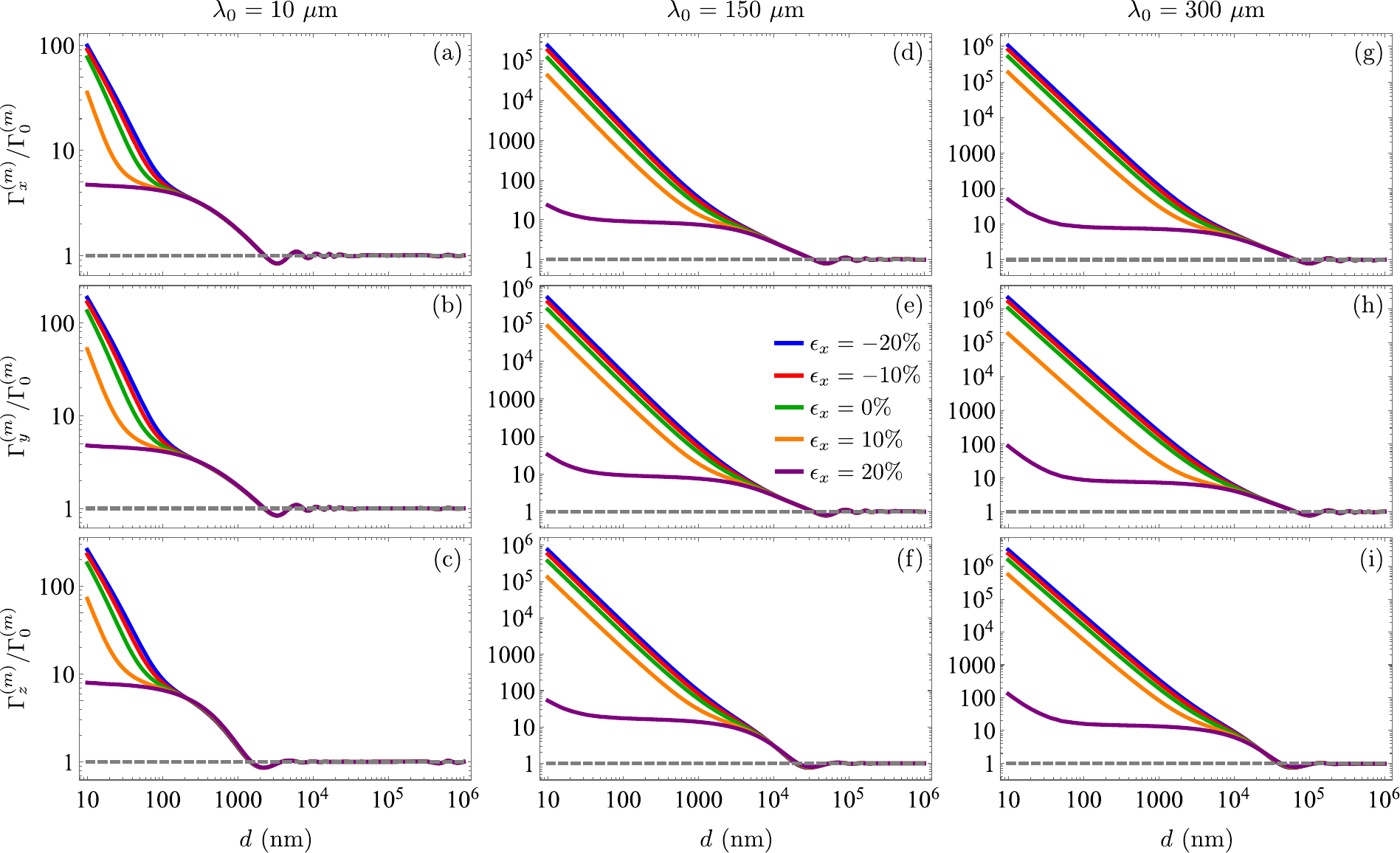}    
	\caption{Magnetic dipole Purcell factors $\Gamma^{(m)}_x/\Gamma_0^{(m)}$, $\Gamma^{(m)}_y/\Gamma_0^{(m)}$, and $\Gamma^{(m)}_z/\Gamma_0^{(m)}$ as functions of the distance $d$ between the emitter and the phosphorene/SiC medium. Here, we considered the uniaxial strain in the $x$ direction with intensities $\epsilon_x=-20, -10, 0, 10, 20 \%$. The emitter's transition wavelengths are (a)-(c) $10$ $\mu$m, (d)-(f) $150$ $\mu$m, and (g)-(i) $300$ $\mu$m.}
	\label{FigPF-Mag-Sx}
\end{figure*}

In Figs.~\ref{FigPF-Ele-Sy} and \ref{FigPF-Mag-Sy}, we presented the electric and magnetic PFs as functions of the separation between the emitter and the phosphorene/SiC half-space for different values of uniaxial strain applied along the $y$ direction. Figures~\ref{FigPF-Ele-Sx} and \ref{FigPF-Mag-Sx} contain the results for the electric and magnetic PFs, respectively, when considering the uniaxial strain applied along the $x$ direction. 

\begin{figure}[t!]
	\centering
	 \includegraphics[width=0.75\linewidth]{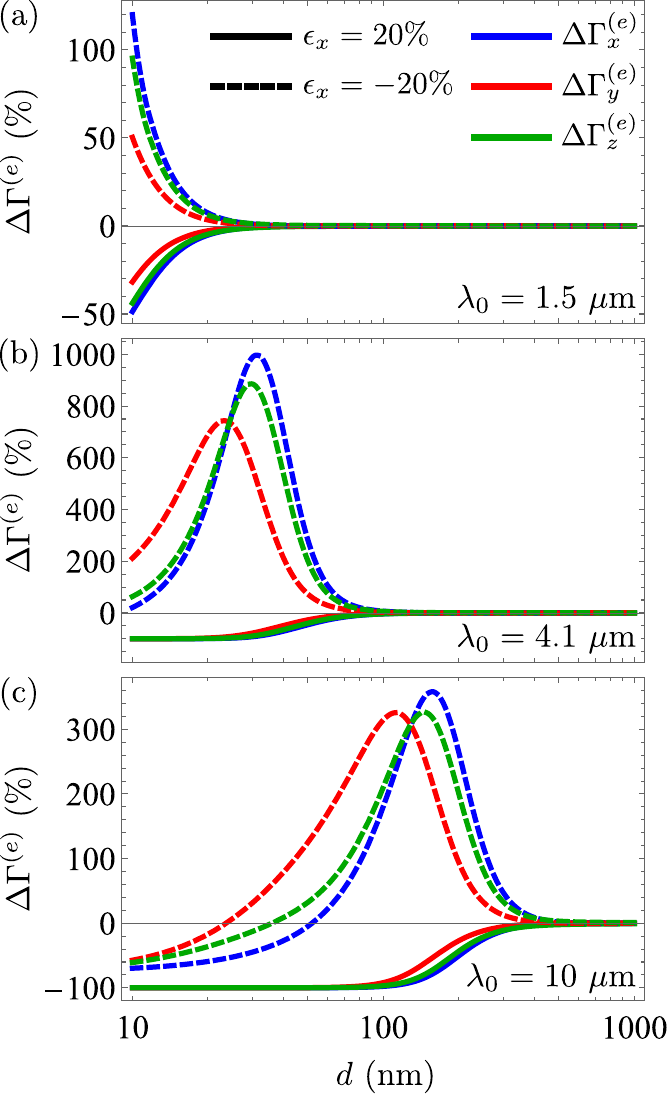}    
	\caption{Percentage variation in electric PFs generated by the uniaxial strain along the $y$ direction as a function of the distance from the emitter to the phosphorene/SiC medium. Solid (dashed) curves show results for $\epsilon_x = 20\% (-20\%)$.}
	\label{FigDeltaPF-Ele-Sx}
\end{figure}

\begin{figure}[t!]
	\centering
	\includegraphics[width=0.735\linewidth]{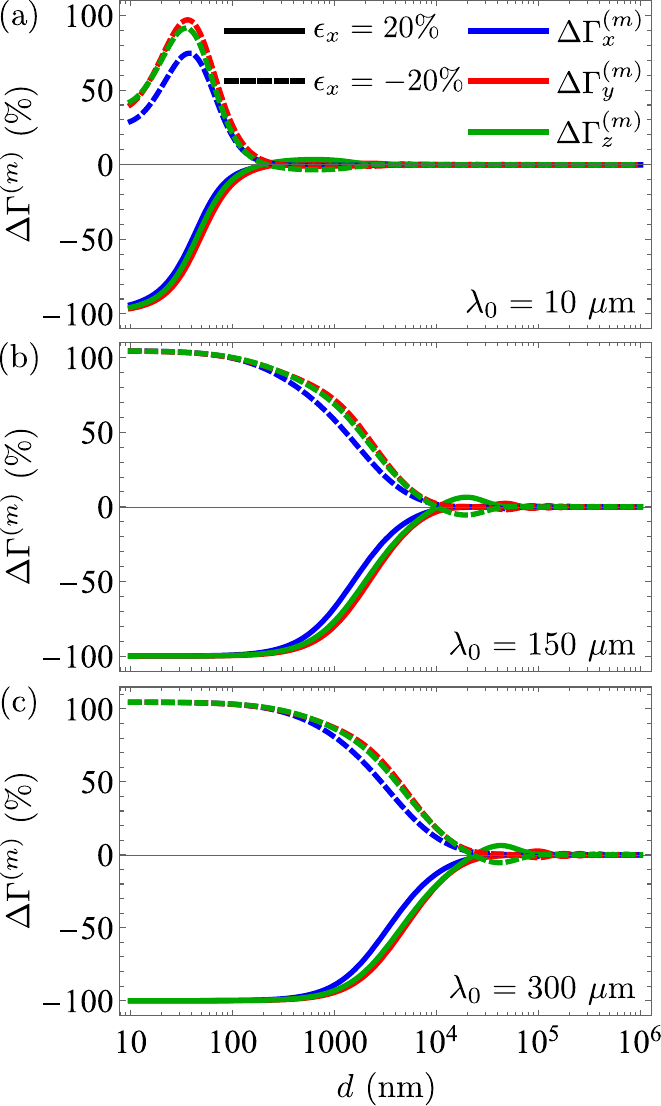}    
	\caption{Percentage variation in magnetic PFs generated by the uniaxial strain along the $x$ direction as a function of the distance from the emitter to the phosphorene/SiC medium. Solid (dashed) curves show results for $\epsilon_x = 20\% (-20\%)$.}
	\label{FigDeltaPF-Mag-Sx}
\end{figure}

Figures~\ref{FigDeltaPF-Ele-Sx} and \ref{FigDeltaPF-Mag-Sx} present the percentage variation in the electric and magnetic PFs, respectively, generated by the uniaxial strain along the $x$ direction as functions of the distance from the emitter to the phosphorene/SiC medium. In both results, the compressive strain may strongly increase the PFs, while the tensile strain nearly suppresses them. We highlight the electric PF for $\lambda_0 = 4.1$ $\mu$m that can be enhanced up to almost $1000 \%$.


\bibliographystyle{apsrev}

\begin{thebibliography}{1}

\bibitem{Purcell1946} E. M. Purcell, H. C. Torrey, and R. V. Pound, Phys. Rev. {\bf 69}, 37 (1946).

\bibitem{Ye-Bizarri-Scintillator-2022} W. Ye, G. Bizarri, M. D. Birowosuto, and L. J. Wong, ACS Photonics {\bf 9}, 3917 (2022).
\bibitem{Kim-Jung-Park-2021} S. K. Kim, S. W. Jung, H.-U. Park, R. Lampande, J. H. Kwon, Org. Electron. {\bf 95}, 106192 (2021).
\bibitem{Huang-Chen-Yang-OptExpress-2022} S. Huang, Y. Chen, Y. Yang, and W. E. I. Sha, Opt. Express {\bf 14}, 24544 (2022).
\bibitem{Kaupp-Hunger-PRApplied-2016} H. Kaupp, T. Hummer, M. Mader, B. Schlederer, J. Benedikter, P. Haeusser, H.-C. Chang, H. Fedder, T. W. Hansch, and D. Hunger, Phys. Rev. Appl. {\bf 6}, 054010 (2016).
\bibitem{Jeantet-Voisin-PRL-2016} A. Jeantet, Y. Chassagneux, C. Raynaud, Ph. Roussignol, J. S. Lauret, B. Besga, J. Esteve, J. Reichel, and C. Voisin, Phys. Rev. Lett. {\bf 116}, 247402 (2016).

\bibitem{Blanco-GarciadeAbajo-2004} L. A. Blanco and F. J. Garc\'{i}a de Abajo, Phys. Rev. B {\bf 69}, 205414 (2004). 
\bibitem{Rosa-Farina-2008} F. S. S. Rosa, T. N. C. Mendes, A. Ten\'{o}rio, and C. Farina, Phys. Rev. A {\bf 78}, 012105 (2008).
\bibitem{Biehs-Greffet-2011} S.-A. Biehs and J.-J. Greffet, Phys. Rev. A  {\bf 84}, 052902 (2011).
\bibitem{Vladimirova-adkov-2012} Y. V. Vladimirova, V. V. Klimov, V. M. Pastukhov, and V. N. Zadkov, Phys. Rev. A {\bf 85}, 053408 (2012).
\bibitem{Kort-Kamp-Farina-2013} W. J. M. Kort-Kamp, F. S. S. Rosa, F. A. Pinheiro, and C. Farina, Phys. Rev. A {\bf 87}, 023837 (2013). 

\bibitem{Klimov-ACSnano-2015} Y.-S. Park, S. Guo, N. S. Makarov, and V. I. Klimov, ACS Nano {\bf 9}, 10386 (2015).
\bibitem{Klimov-NatMat-2019} Y.-S. Park, J. Lim and, V. I. Klimov, Nat. Mater. {\bf 18}, 249 (2019).
\bibitem{Lodahl-Nature-2004} P. Lodahl, A. Floris van Driel, I. S. Nikolaev, A. Irman, K. Overgaag, D. Vanmaekelbergh, and W. L. Vos, Nature {\bf 430}, 654 (2004).
\bibitem{Lodahl-RPM-2015} P. Lodahl, S. Mahmoodian, and S. Stobbe, Rev. Mod. Phys. {\bf 87}, 347 (2015).
\bibitem{vanDriel-PRL-2005} A. F. van Driel, G. Allan, C. Delerue, P. Lodahl, W. L. Vos, and D. Vanmaekelbergh, Phys. Rev. Lett. {\bf 95}, 236804 (2005).

\bibitem{Novotny-book} L. Novotny and B. Hecht, {\it Principles of Nano-Optics}, 2nd ed. (Cambridge University Press, Cambridge, 2006).

\bibitem{Review-QuantumDots} H. Lu, G. M. Carroll, N. R. Neale, and M. C. Beard, ACS Nano {\bf 13}, 939 (2019).

\bibitem{Hussein-Neshev-OLett-2015} R. Hussain, S. S. Kruk, C. E. Bonner, M. A. Noginov, I. Staude, Y. S. Kivshar, N. Noginova, and D. N. Neshev, Opt. Lett. {\bf 40}, 1659 (2015).  
\bibitem{THz-Mag-Purcell} H.-W. Wu, Y. Li, H.-J. Chen, Z.-Q. Sheng, H. Jing, R.-H. Fan, and R.-W. Peng, ACS Appl. Nano Mater. {\bf 2}, 1045 (2019).
\bibitem{Alu-magneticSE-Review} D. G. Baranov, R. S. Savelev, S. V. Li, A. E. Krasnok, and A. Al\'{u}, Laser Photonics Rev. {\bf 11}, 1600268 (2017).
\bibitem{Feng-QD-Lambda-500nm} T. Feng, W. Zhang, Z. Liang, Y. Xu, and E. Miroshnichenko, ACS Photonics {\bf 5}, 678 (2018).

\bibitem{MPF-Silicon-Nanostructures} Y. Brule, P. Wiecha, A. Cuche, V. Paillard, and G. C. des Francs, Opt. Express {\bf 12}, 20360 (2022).
\bibitem{Ferreira-Peres-EPL-2019} B. A. Ferreira and N. M. R. Peres, Europhys. Lett. {\bf 127}, 37002 (2019).
\bibitem{Alaeian-Dionne-PRB-2015} H. Alaeian and J. A. Dionne, Phys. Rev. B {\bf 91}, 245108 (2015). 

\bibitem{Cysne-CasimirPolder-PRA-2014} T. Cysne, W. J. M. Kort-Kamp, D. Oliver, F. A. Pinheiro, F. S. S. Rosa, and C. Farina, Phys. Rev. A {\bf 90}, 052511 (2014).
\bibitem{Silvestre-QR-PRA-2019} M. Silvestre, T. P. Cysne, D. Szilard, F. A. Pinheiro, and C. Farina, Phys. Rev. A {\bf 100}, 033605 (2019).
\bibitem{Abrantes-QR-PRB-2021} P. P. Abrantes, Tarik P. Cysne, D. Szilard, F. S. S. Rosa, F. A. Pinheiro, and C. Farina, Phys. Rev. B {\bf 104}, 075409 (2021).
\bibitem{Rodriguez-Lopez-NatCommun-2017} P. Rodriguez-Lopez, W. J. M. Kort-Kamp, D. A. R. Dalvit, and L. M. Woods, Nat. Commun. {\bf 8}, 14699 (2017). 

\bibitem{Muniz-Farina-Kort-Kamp-2021} Y. Muniz, C. Farina, and W. J. M. Kort-Kamp, Phys. Rev. Res. {\bf 3}, 023061 (2021).
\bibitem{Kort-Kamp-Amorim-FresnelCoefficients} W. J. M. Kort-Kamp, B. Amorim, G. Bastos, F. A. Pinheiro, F. S. S. Rosa, N. M. R. Peres, and C. Farina, Phys. Rev. B {\bf 92}, 205415 (2015).

\bibitem{NFRHTGraphene-PRApp} H. Wu, Y. Huang, L. Cui, and K. Zhu, Phys. Rev. Appl. {\bf 11}, 054020 (2019).
\bibitem{NFRHTGraphene-PRB} L. Ge, K. Gong, Y. Cang, Y. Luo, X. Shi, and Y. Wu, Phys. Rev. B {\bf 100}, 035414 (2019).
\bibitem{Kort-Kamp-PRL-2017} W. J. M. Kort-Kamp, Phys. Rev. Lett. {\bf 119}, 147401 (2017).
\bibitem{MShah-JPD-AppPhys} M. Shah, J. Phys. D: Appl. Phys. {\bf 55}, 105105 (2022).
\bibitem{Abrantes-RET-2021} P. P. Abrantes, G. Bastos, D. Szilard, C. Farina, and F. S. S. Rosa, Phys. Rev. B {\bf 103}, 174421 (2021).

\bibitem{Low-Martin-Moreno-NatureMaterials} T. Low, A. Chaves, J. D. Caldwell, A. Kumar, N. X. Fang, P. Avouris, T. F. Heinz, F. Guinea, L. Martin-Moreno, and F. Koppens, Nat. Mater. {\bf 16}, 182 (2017).
\bibitem{Reserbat-Plantey-ACSPhotonics-2021} A. Reserbat-Plantey, I. Epstein, L. Torre, A. T. Costa, P. A. D. Gonçalves, N. Asger Mortensen, M. Polini, J. C. W. Song, N. M. R. Peres, and F. H. L. Koppens, ACS Photonics {\bf 8}, 85 (2021).
\bibitem{Liu-Mohideen-PRL-2021} M. Liu, Y. Zhang, G.L. Klimchitskaya, V.M. Mostepanenko, and U. Mohideen, Phys. Rev. Lett. {\bf 126}, 206802 (2021).
\bibitem{Guest-NanoLett-2018} C. Husko, J. Kang, G. Moille, J. D. Wood, Z. Han, D. Gosztola, X. Ma, S. Combrie, A. De Rossi, M. C. Hersam, X. Checoury, and J. R. Guest, Nano Lett. {\bf 18}, 6515 (2018).


\bibitem{Phosphorene-First-Syntesis} L. Li, Y. Yu, G. Jun Ye, Q. Ge, X. Ou, H. Wu, D. Feng, X. H. Chen, and Y. Zhang, Nat. Nanotechnol. {\bf 9}, 372 (2014).
\bibitem{Phosphorene-Second-Syntesis} H. Liu, A. T. Neal, Z. Zhu, D. Tomanek, and P. D. Ye, ACS Nano {\bf 8}, 4033 (2014).

\bibitem{Review-Light-Matter-Phosphorene} J. Lu, J. Yang, A. Carvalho, H. Liu, Y. Lu, and C. H. Sow, Acc. Chem. Res. {\bf 49}, 1806 (2016).
\bibitem{Rudenko-Katsnelson-Ph-NoStrain} A. N. Rudenko and M. I. Katsnelson, Phys. Rev. B {\bf 89}, 201408 (2014).
\bibitem{Rodin-Carvalho-CastroNeto-Ph-NoStrain} A. S. Rodin, A. Carvalho, and A. H. Castro Neto, Phys. Rev. Lett. {\bf 112}, 176801 (2014).
\bibitem{Casimir-Torque-Phosphorene} P. Thiyam, P. Parashar, K. V. Shajesh, O. I. Malyi, M. Bostrom, K. A. Milton, I. Brevik, and C. Persson, Phys. Rev. Lett. {\bf 120}, 131601 (2018).

\bibitem{Phosphorene-SE-Twisting} H. Mu, T. Wang, D. Zhang, W. Liu, T. Yu, and Q. Liu, Opt. Express {\bf 2}, 1037 (2021).
\bibitem{Phosphorene-SE-NLayer} B. Sikder, S. H. Mayem, and S. Z. Uddin, Opt. Express {\bf 26}, 47152 (2022).
\bibitem{Phosphorene-SE-PRAplied-2019} E. van Veen, A. Nemilentsau, A. Kumar, R. Roldan, M. I. Katsnelson, T. Low, and S. Yuan, Phys. Rev. Appl. {\bf 12}, 014011 (2019).
\bibitem{Phosphorene-SE-Bilayer} L. Sun, G. Zhang, S. Zhang, and J. Ji, Opt. Express {\bf 13}, 14270 (2017).

\bibitem{Peeters-StrainPhosphorene-Model} E. Taghizadeh Sisakht, F. Fazileh, M. H. Zare, M. Zarenia, and F. M. Peeters, Phys. Rev. B {\bf 94}, 085417 (2016).
\bibitem{Midtvedt-Lewenkopf-Croy-2DMat} D. Midtvedt, C. H. Lewenkopf, and A. Croy, 2D Mater. {\bf 3}, 011005 (2016).
\bibitem{Midtvedt-Lewenkopf-Croy-JPCM} D. Midtvedt, C. H. Lewenkopf, and A. Croy, J. Phys.: Condens. Matter {\bf 29}, 185702 (2017).

\bibitem{Flexibility-Phosphorene-1} Q. Wei and X. Peng, Appl. Phys. Lett. {\bf 104}, 251915 (2014).
\bibitem{Flexibility-Phosphorene-2} X. Peng, Q. Wei, and A. Copple, Phys. Rev. B {\bf 90}, 085402 (2014).

\bibitem{Exp-Strain-Ph-1} S. Huang, G. Zhang, F. Fan, C. Song, F. Wang, Q. Xing, C. Wang, H. Wu, and H. Yan, Nat. Commun. {\bf 10}, 2447 (2019).
\bibitem{Exp-Strain-Ph-2} J. Quereda, P. San-Jose, V. Parente, L. Vaquero-Garzon, A. J. Molina-Mendoza, N. Agrait, G. Rubio-Bollinger, F. Guinea, R. Roldan, and A. Castellanos-Gomez, Nano Lett. {\bf 16}, 2931 (2016). 

\bibitem{Alidoust-Akola-PRB-2021} M. Alidoust, E. E. Isachsen, K. Halterman, and J. Akola, Phys. Rev. B {\bf 104}, 115144 (2021). 
\bibitem{Yan-Zhang-Wang-Zhang-Optcond} C. H. Yang, J. Y. Zhang, G. X. Wang, and C. Zhang, Phys. Rev. B {\bf 97}, 245408 (2018).
\bibitem{Li-Peeters-2018} L. L. Li and F. M. Peeters, Phys. Rev. B {\bf 97}, 075414 (2018).
\bibitem{Li-Peeters-2017} L. L. Li, D. Moldovan, P. Vasilopoulos, and F. M. Peeters, Phys. Rev. B {\bf 95}, 205426 (2017).
\bibitem{Faria-Junior-Low-energyModel} P. E. Faria Junior, M. Kurpas, M. Gmitra, and J. Fabian, Phys. Rev. B {\bf 100}, 115203 (2019). 
\bibitem{Low-Rodin-Optcond} T. Low, A. S. Rodin, A. Carvalho, Y. Jiang, H. Wang, F. Xia, A. H. Castro Neto, Phys. Rev. B {\bf 90}, 075434 (2014).

\bibitem{Non-Local_x_Local-PF-Abajo} R. Petersen, T. G. Pedersen, and F. Javier Garc\'ia de Abajo, Phys. Rev. B {\bf 96}, 205430 (2017).

\bibitem{Szilard_SE_VO2} D. Szilard, W. J. M. Kort-Kamp, F. S. S. Rosa, F. A. Pinheiro, and C. Farina, J. Opt. Soc. Am. B {\bf 36}, C46 (2019).

\bibitem{Purcell-Phosphorene-OptExpress-Sikder} B. Sikder, S. Hasan Nayem, and S. Zia Uddin, Opt. Express {\bf 26}, 47152 (2022).

\bibitem{Das-Roelofs-ACS-Nano-2014} S. Das, M. Demarteau, and A. Roelofs, ACS Nano {\bf 8}, 11730 (2014).

\bibitem{Hulet-Kleppner-dipoleTHz-PRL-1985} R. G. Hulet, E. S. Hilfer, and D. Kleppner, Phys. Rev. Lett. {\bf 55}, 2137 (1985).


\bibitem{Gaudreau-Koppens-NanoLett-2013} L. Gaudreau, K. J. Tielrooij, G. E. D. K. Prawiroatmodjo, J. Osmond, F. J. García de Abajo, and F. H. L. Koppens, Nano Lett. {\bf 13}, 2030 (2013).

\bibitem{Szilard-PRB-2016} D. Szilard, W. J. M. Kort-Kamp, F. S. S. Rosa, F. A. Pinheiro, and C. Farina, Phys. Rev. B {\bf 94}, 134204 (2016).

\bibitem{HarrisonBook} W. A. Harrison, {\it Elementary Electronic Structure} (World Scientific, Singapore, 1999).

\bibitem{Novko-Opticond-Phosphorene} D. Novko, K. Lyon, D. J. Mowbray, and V. Despoja, Phys. Rev. B {\bf 104}, 115421 (2021).
\bibitem{Moshayedi-Opticond-Phosphorene} M. Moshayedi, M. R. Preciado Rivas, and Z. L. Miskovic, Phys. Rev. B {\bf 105}, 075429 (2022).

\bibitem{Cysne-Rappoport-DisorderGraphene} T. P. Cysne, T. G. Rappoport, A. Ferreira, J. M. Viana Parente Lopes, and N. M. R. Peres, Phys. Rev. B {\bf 94}, 235405 (2016). 

\bibitem{Zhu-Zhang-Li-Relaxation-Time} L. Zhu, G. Zhang, and B. Li, Phys. Rev. B {\bf 90}, 214302 (2014).
\bibitem{Lv-Lu-Relaxation-Time} H. Y. Lv, W. J. Lu, D. F. Shao, and Y. P. Sun, Phys. Rev. B {\bf 90}, 085433 (2014).

\bibitem{Guinea-Martin-Moreno-PRR2019} Tetiana M. Slipchenko, Jurgen Schiefele, Francisco Guinea, and Luis Martin-Moreno, Phys. Rev. Res. {\bf 1}, 033049 (2019).

\bibitem{DL-SiC-PaliK_book} E. W. Palik, {\it Handbook of Optical Constants of Solids} (Academic Press, San Diego, 1985).
\bibitem{Moreno-FresnelCoefficients} M. Moreno, Phys. Rev. A {\bf 93}, 013832 (2016).
 

\end{thebibliography}

\end{document}